\documentclass[a4paper,11pt]{article}
\pdfoutput=1 

\usepackage{jheppub} 

\usepackage{amsmath}
\usepackage{amssymb}
\usepackage{young}
\usepackage[vcentermath]{youngtab}
\usepackage{dsfont}
\usepackage{braket}
\usepackage{comment}
\usepackage{textgreek} 
\usepackage{simplewick}
\allowdisplaybreaks
\usepackage{mathrsfs}
\usepackage[mathscr]{euscript}
\usepackage{tikz}
\usepackage{tensor}
\usepackage{mathtools}

\numberwithin{equation}{section}

\def\be{\begin{equation}}
\def\ee{\end{equation}}
\newcommand{\tr}{\text{tr}\,}

\def \nn {\nonumber}
\def \la {\langle}
\def \ra {\rangle}
\def \SL {\mathrm{SL}}
\def \C {\mathds{C}}
\def \R {\mathds{R}}
\def \I {\mathbb{I}}
\def \contact {\mathcal{C}}
\def \d {\mathrm{d}}

\newcommand*{\tran}{^{\mkern-1.5mu\mathsf{T}}}

\title{Scattering Equations in AdS: \\Scalar Correlators in Arbitrary Dimensions}

\author{Lorenz Eberhardt,}
\author{Shota Komatsu,}
\author{Sebastian Mizera}
\affiliation{School of Natural Sciences, Institute for Advanced Study, \\
\hspace*{0.3cm}1 Einstein Drive, Princeton, NJ 08540, USA}
\emailAdd{elorenz@ias.edu}
\emailAdd{skomatsu@ias.edu}
\emailAdd{smizera@ias.edu}

\abstract{We introduce a bosonic ambitwistor string theory in AdS space.  Even though the theory is anomalous at the quantum level, one can nevertheless use it in the classical limit to derive a novel formula for correlation functions of boundary CFT operators in arbitrary space-time dimensions. The resulting construction can be treated as a natural extension of the CHY formalism for the flat-space S-matrix, as it similarly expresses tree-level amplitudes in AdS as integrals over the moduli space of Riemann spheres with punctures. These integrals localize on an operator-valued version of scattering equations, which we derive directly from the ambitwistor string action on a coset manifold. As a testing ground for this formalism we focus on the simplest case of ambitwistor string coupled to two current algebras, which gives bi-adjoint scalar correlators in AdS. In order to evaluate them directly, we make use of a series of contour deformations on the moduli space of punctured Riemann spheres and check that the result agrees with tree level Witten diagram computations to all multiplicity. We also initiate the study of eigenfunctions of scattering equations in AdS, which interpolate between conformal partial waves in different OPE channels, and point out a connection to an elliptic deformation of the Calogero-Sutherland model.}

\begin{document}
	
\maketitle
\flushbottom
\setcounter{page}{2}

\section{Introduction}

In 2003 Witten discovered that flat-space scattering amplitudes of ${\cal N}=4$ super Yang-Mills can be understood in terms of a string theory in twistor space \cite{Witten:2003nn}. In this reformulation the amplitudes are computed as certain localization integrals involving the moduli space of Riemann surfaces with $n$ punctures, ${\cal M}_{0,n}$ \cite{Witten:2003nn,Roiban:2004yf}. We now understand that this discovery was not merely a fluke; not only does it generalize to gravitational theories \cite{Berkovits:2004jj,Cachazo:2012da,Cachazo:2012kg}, but also seems to have non-supersymmetric and higher-dimensional counterparts. This generalization is known as the \emph{ambitwistor} string theory, first introduced by Mason and Skinner \cite{Mason:2013sva} in 2013 (see \cite{Adamo:2013tsa,Geyer:2014fka,Casali:2015vta,Geyer:2015bja,Geyer:2015jch,Geyer:2016wjx,Roehrig:2017gbt,Geyer:2017ela,Geyer:2018xwu,Geyer:2019hnn} for follow-up work). It successfully explained the physical origin of the Cachazo-He-Yuan (CHY) formulae \cite{Cachazo:2013hca,Cachazo:2013iea} computing scattering amplitudes in various scalar, gauge, and gravity theories in terms of localization integrals on ${\cal M}_{0,n}$ on the support of the so-called \emph{scattering equations} \cite{Cachazo:2013gna}. To be a bit more precise, in this formalism scattering amplitudes are computed via integrals of the form
\be
\int_{{\cal M}_{0,n}} \!\!\! {\cal I}_\text{L}\, {\cal I}_\text{R} \prod_{i} \d z_i\, \bar{\delta}\bigg(\sum_{j \neq i}\frac{2p_i \cdot p_j}{z_i - z_j}\bigg) \underbrace{\int_{\text{Mink}_{d+1}} \!\!\!\!\!\! \d x\, \prod_{i=1}^{n} e^{i x {\cdot} p_i}}_{(2\pi)^{d+1}\delta^{d+1} (\sum_{i=1}^{n} p_i)}\,,
\ee
where ${\cal I}_\text{L}$ and ${\cal I}_\text{R}$ are functions of external data and the positions $z_i$ of punctures on the worldsheet that differ depending on the specific matter content of the theory. The (anti-holomorphic) delta functions localize on the solution of scattering equations. The final term is simply a delta function imposing momentum conservation, which, foreshadowing the results of this paper, we suggestively wrote as a trivial integral of $n$ plane waves interacting at a single point $x$ in $(d{+}1)$-dimensional Minkowski space.

Recent years have seen a surge of interest in further extending the ambitwistor formulation to curved spaces, in particular to plane wave backgrounds \cite{Adamo:2014wea,Adamo:2017sze,Adamo:2018hzd,Adamo:2018ege} (see also \cite{Adamo:2012nn,Adamo:2013tja,Adamo:2013tca,Adamo:2015ina} for twistor models). It then seems rather prudent to ask: what about one of the simplest curved backgrounds of physical interest, the anti-de Sitter (AdS) space?

The AdS space-time has several distinctive features that are not shared by the flat space or any other curved backgrounds. Firstly, it provides a symmetry-preserving infrared regulator of the flat-space physics. By analyzing theories in AdS and carefully taking the flat-space limit, one can study the infrared dynamics in flat space without needing to deal with infrared divergences. This was pointed out initially by Callan and Wilczek \cite{Callan:1989em}, and the idea was developed further in \cite{Aharony:2012jf}, who studied Yang-Mills theory in AdS with the aim of understanding confinement. Secondly, the AdS/CFT correspondence \cite{Maldacena:1997re} relates the amplitudes in AdS to the correlation functions in the dual conformal field theory (CFT), providing powerful tools to study the latter. For instance, the tree-level amplitudes of supergravity in AdS compute the correlation functions of strongly-coupled large $N$ superconformal field theories, which are otherwise difficult to analyze.
Thirdly, the correlation functions of operators on the boundary of AdS can be studied using the conformal symmetry SO$(2,d)$ and the techniques of the conformal bootstrap \cite{Rattazzi:2008pe,Poland:2018epd}. In particular, starting from the seminal works \cite{Rastelli:2016nze,Aharony:2016dwx}, four-point functions in various supergravity theories in AdS were computed at tree- and loop-levels  by analytically solving the conformal crossing equation\footnote{See \cite{Alday:2020lbp,Alday:2020dtb} for the most recent progress in this direction.}. However, extending such analyses to higher-point functions seems much harder owing to the complexity of the relevant conformal blocks, although important progress has been made recently both for the conformal blocks \cite{Rosenhaus:2018zqn,Parikh:2019ygo,Jepsen:2019svc,Parikh:2019dvm,Fortin:2019zkm,Fortin:2020yjz,Fortin:2020bfq,Hoback:2020pgj} and the analysis of the crossing equation \cite{Goncalves:2019znr}. Given this situation, it would be useful to develop alternative approaches that work equally well for higher-point functions.

In this work we initiate the study of ambitwistor string theory in AdS space, whose correlation functions compute the conformal field theory correlators, as well as the generalization of the scattering equations this formulation entails. As a starting point we consider perhaps the simplest setup corresponding to a bosonic version of string theory with scalar vertex operators. To this end, we propose an ambitwistor action on the coset space
\be \label{eq:intro coset}
\mathrm{AdS}_{d+1} =\frac{\mathrm{SO}(2,d)}{\mathrm{SO}(1,d)}\, .
\ee
As will be discussed in detail later, the theory is inconsistent at quantum level owing to various anomalies. Nevertheless we show that it leads to a well-defined CHY-like formula in the infinite tension limit; namely in the limit in which the worldsheet theory becomes classical and anomalies become irrelevant.

Making use of the AdS embedding formalism \cite{Costa:2011mg} we find that correlators take the general form
\be\label{eq:intro}
\int_{{\cal M}_{0,n}} {\cal I}_\text{L}\, {\cal I}_\text{R} \prod_{i} \d z_i\, \bar{\delta}\bigg(\sum_{j \neq i}\frac{2D_i \cdot D_j}{z_i - z_j}\bigg) \underbrace{\int_{\text{AdS}_{d+1}} \!\!\! \d X \prod_{i=1}^{n}\frac{1}{(-2X\cdot P_{i})^{d}}}_{\mathcal{C}(P_1,\dots,P_n)}\, ,
\ee
where, once again, ${\cal I}_L$ and ${\cal I}_R$ depend on the specific matter content. Here $P_i^A$ are the embedding space coordinates and $D_i^{AB} = P_i^A \partial_{P_{i,B}} - P_i^B \partial_{P_{i,A}}$ are the scalar conformal generators (the notation is reviewed in Section~\ref{sec:embedding-space}). This expression acts on the scalar contact diagram $\mathcal{C}$ of $n$ bulk-to-boundary propagators meeting at the point $X$, which replaces the flat-space momentum conservation delta function. In the case of bi-adjoint scalar, which we mainly focus on, the integrands ${\cal I}_\text{L}$ and ${\cal I}_\text{R}$ are given by Parke-Taylor factors known from \cite{Witten:2003nn,Cachazo:2013hca}.

The starting point of our analysis is the action of the bosonic ambitwistor string on a group manifold $\text{G}$. Viewing $\text{Mink}_{d+1}$ as an abelian group, this constitutes a non-abelian generalization of the standard ambitwistor action. It has a manifest $\text{G}_\text{L}\times \text{G}_\text{R}$ symmetry, corresponding to left- and right-multiplication on the group. In order to pass to the coset, we gauge a subgroup $\mathrm{H} \subset \text{G}_\text{R}$. Consequently, the BRST operator of the ambitwistor string receives a further term that implements this gauging. 
Specializing to the coset \eqref{eq:intro coset}, we construct vertex operators and compute their correlation functions, which leads to the CHY-like formula \eqref{eq:intro}. 

The main novelty compared to flat space is the fact that the AdS scattering equations are operator-valued. This inhibits direct computations, as nothing about the positions (or even the number) of localization points can be assumed. In order to alleviate this problem we evaluate correlators by first writing the expression \eqref{eq:intro} as a middle-dimensional contour integral, followed by a series of contour deformations that, following \cite{Dolan:2013isa}, localize the correlator on the worldsheet configurations factorizing into trivalent graphs. These are essentially in one-to-one map with trivalent Witten diagrams. Using this strategy we indeed demonstrate that the result of AdS bi-adjoint scalar correlation functions agrees with a Witten diagram computation to arbitrary multiplicity $n$.

A possible alternative for evaluating \eqref{eq:intro} is to decompose the integrand into joint eigenfunctions of the scattering equations. This would allow us to replace the delta functions with operators in the arguments with standard delta functions, and compute the integral explicitly. In this paper, we take an initial step in this direction: We study the eigenvalue equations of the scattering equation for the four-point functions and demonstrate an interesting connection to quantum integrable models:  After a change of variables to the so-called ``pillow coordinates'' introduced by Zamolodchikov \cite{zamolodchikov1987conformal} in the analysis of 2d CFT, the eigenvalue equations coincide with the Schr{\"o}dinger equation of the $BC_2$ Inozemtsev model, which is an elliptic deformation of the $BC_2$ Calogero-Sutherland model. To our knowledge, this is the first example in which the pillow coordinates show up naturally in the analysis of higher-dimensional CFTs. We also emphasize that the eigenfunctions of the AdS scattering equations are interesting objects by themselves since they interpolate between conformal partial waves in different OPE channels. 

Finally, let us comment on the relation to the unpublished work of Roehrig and Skinner \cite{SkinnerTalk,Roehrig:2019zqb}, by which this work was inspired. They studied ambitwistor string theory on the group manifold $\mathrm{AdS}_3 \times \mathrm{S}^3 \cong \mathrm{SL}(2,\R) \times \mathrm{SU}(2)$. The key difference to our work is that we consider the model on a \emph{coset space} and in fact that is what allows us to work in arbitrary space-time dimension $d$. In addition, since $\mathrm{AdS}_3 \times \mathrm{S}^3$ is a well-defined supergravity background, their theory is free of anomalies and is consistent at quantum level. Despite these differences, Roehrig and Skinner arrived at a formula similar to \eqref{eq:intro} with the appropriate contact diagram $\mathcal{C}$ on $\mathrm{AdS}_3 \times \mathrm{S}^3$ and attempted to evaluate it in Mellin space, as opposed to contour deformations employed here.

\paragraph{Outline.} The paper is organized as follows. In Section~\ref{sec:bosonic-coset} we study the bosonic ambitwistor string action with various targets. After a review of the flat-space case, we provide generalizations to group manifolds and coset spaces and analyze them on the classical and quantum levels. In Section~\ref{sec:bosonic-AdS} we focus on the case of $\mathrm{AdS}_{d+1}$ and construct scalar vertex operators based on a pair of internal current algebras, as well as demonstrate localization of correlation functions on the AdS scattering equations. In Section~\ref{sec:correlators} we compute the correlators using contour manipulations, thus proving a recursion relation for bi-adjoint scalar correlation functions in AdS. We then demonstrate their equivalance to the Witten diagram computation. In Section~\ref{sec:eigenfunctions} we study eigenfunctions of scattering equations in AdS and explain a connection to the $BC_2$ Inozemtsev model, which is an elliptic deformation of the $BC_2$ Calogero-Sutherland model.  Finally, in Section~\ref{sec:discussion} we give a discussion of the results and future directions. Appendix~\ref{sec:notation} contains a summary of the notation.

\paragraph{Note added.} After completion of this work, we became aware that Roehrig and Skinner were writing up and planning to publish their results \cite{Roehrig:2020kck}. We therefore decided to coordinate a simultaneous release of our papers on arXiv. We thank them for kindly agreeing to do so.

\section{\label{sec:bosonic-coset}Bosonic ambitwistor string on a coset manifold}

In this section, we will start to build up the ambitwistor string on a general coset manifold. We will restrict ourselves to the bosonic ambitwistor string, since this is the relevant ambitwistor string for the bi-adjoint scalar theory and is technically simpler. We start with a very brief review of the flat space ambitwistor string, then generalize to a group manifold and finally to a coset manifold. Note that the first generalization was already discussed in \cite{Roehrig:2019zqb} in the case of the type II ambitwistor string, where the goal was to describe supergravity on $\text{AdS}_3$. We should mention that it is actually not obvious that our results agree with the results of  \cite{Roehrig:2019zqb} in the case of $\text{AdS}_3$, since we treat $\text{AdS}_3\times \text{S}^3$ as coset manifold, whereas it was treated as a group manifold in \cite{Roehrig:2019zqb}.

\subsection{Flat space}
We consider the flat-space bosonic ambitwistor string \cite{Mason:2013sva}, whose action takes the form\footnote{We always write small $p$, since capital $P$ will appear later as the embedding coordinate in AdS.}
\begin{align}
S=\frac{1}{2\pi}\int  p_\mu \bar{\partial} X^\mu +S_\text{current}^{(1)}+S_\text{current}^{(2)}\,,
\end{align}
where $S_\text{current}^{(1,2)}$ is the action for two current algebras at level $k$. We will in the following concentrate on the first term in the action. The system has two kinds of symmetries:

\paragraph{Reparametrization invariance.}
Under a holomorphic change of coordinates parametrized by a holomorphic vector field $v$, we have
\begin{subequations} 
\begin{align}
\delta X_\mu &=v \partial X_\mu \,, \\
\delta p_\mu &=\partial(v  p_\mu) \,.
\end{align}
\end{subequations}
The action changes as
\be 
\delta S=\frac{1}{2\pi} \int  \bar{\partial}v p_\mu\partial X^\mu+\delta S_\text{current}^{(1)}+\delta S_\text{current}^{(2)}\,.
\ee
Thus, the action is invariant and the holomorphic energy momentum tensor is
\be  
T=p_\mu \partial X^\mu+T_\text{current} .
\ee
Here and in the following $T_\text{current}$ will denote the energy-momentum tensor of the internal current algebras.

\paragraph{Ambitwistor symmetry.}
The action has an additional symmetry that, when gauged reduces the target space from from $\mathds{C}^d$ to the ambitwistor space $\mathds{A}$,
\be 
\delta X_\mu=\alpha p_\mu\,,
\ee
where $\alpha$ is an arbitrary holomorphic vectorfield on the Riemann surface. The action changes according to
\be 
\delta S=\frac{1}{4\pi}\int \bar{\partial}\alpha p_\mu p^\mu\,,
\ee
and thus the corresponding holomorphic current is $H=p_\mu p^\mu$.

In the ambitwistor string, both of these symmetries are gauged. For more detail on this model we refer the reader to the original paper \cite{Mason:2013sva}.

\subsection{Group manifold}
\paragraph{Classical theory.}
We now promote this to a group manifold, where we have
\begin{align}
S=\frac{1}{2\pi\hbar}\int  \text{tr}\left(p g^{-1}\bar{\partial} g\right)  +S_\text{current}^{(1)}+S_\text{current}^{(2)}\,,
\end{align}
where we replaced $\bar{\partial}X^\mu$ by the anti-holomorphic version of the Maurer-Cartan form or rather its pullback on the worldsheet by the group-valued field $g$. $p$ becomes a Lie-algebra valued $(1,0)$ form in this setting. The trace is an invariant trace on the Lie algebra and is further discussed below. $S_\text{current}^{(1,2)}$ does not change and we again focus on the first term. We also introduced a parameter $\hbar$, which can be interpreted as a ratio between the string length and the size of the group:
\be 
\hbar\sim \frac{\ell_\text{string}}{\ell_\text{Group}}\ .
\ee
Since the correlation functions depend  only on this ratio not on individual lengths, we will in the following set $\ell_\text{Group}=1$, but keep $\hbar$. In the context of AdS, $\ell_\text{Group}$ will become $\ell_\text{AdS}$. 
We gave the name $\hbar$, since it should be viewed as $\hbar$ on the worldsheet and we will call the $\hbar\to 0$ limit  the {\it classical limit} in what follows. We remark that the prefactor of the action does not involve $\ell_\text{string}^{-2}$ as in the physical string, but only $\ell_\text{string}^{-1}$ because of the first order form.

We should note that there is an equivalent action, where we use $\bar{\partial}g g^{-1}$ instead of $g^{-1}\bar{\partial}g$.
The action has a holomorphic $\text{G}_\text{L}(z) \times \text{G}_\text{R}(z)$ symmetry. We can multiply
\begin{subequations}
\begin{align} 
g &\mapsto g_\text{L}(z) g g_\text{R}(z)^{-1}\,, \\
p &\mapsto g_\text{R}(z) p g_\text{R}(z)^{-1}\,.
\end{align}
\end{subequations}
This hence leads to two current algebras in the quantum theory. Start with $g_\text{L}$. We find that the corresponding conserved current is 
\be 
J_\text{L}=g p g^{-1}\,,
\ee
and the equations of motion show indeed that this current is holomorphic.  For the right-multiplication symmetry, the current is
\be 
J_\text{R}=p\,,
\ee
as one can check by direct computation.

The two additional symmetries we discussed for the flat space model are also present and take the form
\begin{subequations}
\begin{align}
\delta g&=v \partial g \,, \\
\delta p &=\partial(v  p) \,.
\end{align}
\end{subequations}
and
\be 
\delta g=\alpha g p\,.
\ee
They lead to the currents $T=\tr(p g^{-1} \partial g)+T_\text{current}$ and $H=\tr(p^2)$, which we again want to gauge.

\paragraph{Quantum theory.}
We next compute the variation of the action with respect to the left-symmetry. This gives for $g_\text{L}=1+\omega_\text{L}$
\begin{align}
\delta_{\omega_\text{L}} S= \frac{1}{\hbar}\int \text{tr}\left(g p g^{-1} \bar{\partial} \omega_\text{L}\right)=  \frac{1}{\hbar}\int \bar{\partial}\,\text{tr}\left(g p g^{-1} \omega_\text{L}\right)=\frac{1}{2\pi i \hbar} \oint \mathrm{d}z\ \omega_\text{L}^a J_\text{L}^a\,.
\end{align}
Here we  introduced a component notation. $a$, $b$, $c$, \dots will in the following denote an adjoint index of the group $\text{G}$. Since we need in the following quite a large number of different indices and fields, we invite the reader to check Appendix~\ref{sec:notation}, where our notation is summarized.
Inserting this into a correlator gives
\be 
\delta_{\omega_\text{L}} \left\langle X \right\rangle=\frac{1}{2\pi i\hbar} \oint \mathrm{d}z \ \omega_\text{L}^a(z) \left\langle J_\text{L}^a (z) X \right\rangle
\ee
On the other hand, we can compute $\delta_{\omega_\text{L}}$ of various quantities directly.  We have
\be 
\delta_{\omega_\text{L}} g=\omega_\text{L} g\,, \qquad \delta_{\omega_\text{L}} J_\text{L}=[\omega_\text{L},J_\text{L}]\,,  \qquad \delta_{\omega_\text{L}} J_\text{R}=0\,.
\ee
Thus, we deduce the OPEs
\begin{subequations}
\begin{align}
J_\text{L}^a(z) J_\text{L}^b(w)&\sim \frac{\hbar\tensor{f}{^{ab}_c} J^c_\text{L}(w)}{z-w}\,, & J_\text{L}^a(z) g(w)&\sim -\frac{\hbar t^a g(w)}{z-w}\,,\\
J_\text{R}^{\bar{a}}(z) J_\text{R}^{\bar{b}}(w)&\sim \frac{\hbar\tensor{f}{^{\bar{a}\bar{b}}_{\bar{c}}} J^{\bar{c}}_\text{R}(w)}{z-w}\,, & J_\text{R}^{\bar{a}}(z) g(w)&\sim \frac{\hbar g(w)t^{\bar{a}}}{z-w}\,, \\
J_\text{R}^{\bar{a}}(z) J_\text{L}^b(w)&\sim 0\,,& g(z) g(w) &\sim 0\,.
\end{align}
\end{subequations}
Here $t^a$ are the generators of the representation in which $g$ transforms. For definiteness, we take it to be the fundamental representation. We have barred indices of $\text{G}_\text{R}$. At this point, $\hbar$ is not important since we could simply rescale the generators to remove it. However, it will play an important role later.
One can similarly derive the other OPEs. Clearly, not all fields are independent.

To continue, let us assume that $\text{G}$ has a biinvariant trace $\tr$, which is the trace we already used in writing down the action. We normalize generators in the fundamental representation such that\footnote{In some cases of interest, like $\mathfrak{psu}(1,1|2)$ or $\mathfrak{psu}(2,2|4)$, there is no fundamental representation, but we can pick any other representation.}
\be 
\tr(t^a t^b)=\delta^{ab}\,.
\ee
For computations, it is also convenient to define
\begin{subequations}
\begin{align} 
j_{\text{L}}(z)&=-\partial g g^{-1}\,, \\
j_{\text{R}}(z)&= g^{-1}\partial g\,.
\end{align}
\end{subequations}
By direct computation, they satisfy the OPE
\begin{align} 
J^a_\text{L}(z) j_{\text{L}}^b(w) &\sim \frac{\hbar \,\delta^{ab}}{(z-w)^2}+\frac{\hbar \tensor{f}{^{ab}_c}j_{\text{L}}^c(w)}{z-w}\,, \\
J^{\bar{a}}_\text{R}(z) j_{\text{R}}^{\bar{b}}(w) &\sim \frac{\hbar \,\delta^{\bar{a}\bar{b}}}{(z-w)^2}+\frac{\hbar\tensor{f}{^{\bar{a}\bar{b}}_{\bar{c}}}j_{\text{R}}^{\bar{c}}(w)}{z-w}\,.
\end{align}
The cross OPEs $J_\text{L}^a(z) j_\text{R}^{\bar{b}}(w)$ involve new fields, but we shall not need them.

To go on, let us assume that the group is simple.
As we showed classically, the energy-momentum tensor of the system is
\begin{align}
T=(J^a_\text{L} j^a_{\text{L}})+T_\text{current}\,,
\end{align}
where $T_\text{current}$ is the Sugawara tensor of the current algebras. We could alternatively also use $J_\text{R}$ to construct the energy-momentum tensor. It turns out that this energy-momentum tensor receives a quantum correction. We find that the unique energy-momentum tensor for which $J_\text{L}$ and $g$ are primaries of conformal weight $1$ and $0$, respectively, takes the form
\be 
T=(J^a_\text{L} j^a_{\text{L}})-\hbar h^\vee_\mathfrak{g} (j^a_\text{L} j^a_{\text{L}})+T_\text{current}\,,
\ee
where $h^\vee_\mathfrak{g}$ is the dual Coxeter number. The fact that the second term is proportional to the dual Coxeter number indicates that it is a quantum correction.

The central charge is
\be 
c_\text{matter}=2\, \text{dim}(\text{G})+c_\text{current}^{(1)}+c_\text{current}^{(2)}\,. \label{eq:group cmatter}
\ee
Next, we look at the current that corresponding to the other classical symmetry that we want to gauge,
\be 
H(z)=(J^a_\text{L}J^a_\text{L})\,.
\ee
Classically, we would have expected that $H(z)$ has regular OPE with itself. 
However, this property is broken at the quantum level and there seems no way to correct $H(z)$ to enforce this property. However, $H(z)$ at least closes on itself, since up to a prefactor it is the Sugawara tensor of the current $J^a_\text{L}$.
These fields satisfy the following gauge algebra:
\begin{subequations}
\begin{align}
T(z)T(w) &\sim \frac{\hbar^2 c_\text{matter}}{2(z-w)^4} +\frac{2\hbar T(w)}{(z-w)^2}+\frac{\hbar \partial T(w)}{z-w}\,, \\
T(z)H(w) &\sim \frac{2\hbar H(w)}{(z-w)^2}+\frac{\hbar \partial H(w)}{z-w}\,,\\
H(z)H(w) &\sim \frac{4\hbar^2 h^\vee_\mathfrak{g} H(w)}{(z-w)^2}+\frac{2\hbar^2 h^\vee_\mathfrak{g} \partial H(w)}{z-w}\, .
\end{align}
\end{subequations}
This is almost the same same gauge algebra as in flat space, up to the correction due to the dual Coxeter number. We can write down a BRST operator implementing these constraints at the quantum level:
\be 
Q=\oint \mathrm{d}z \ \left(c T+\hbar^{-1}\tilde{c}H+2 h^\vee_\mathfrak{g} (\tilde{b}\tilde{c}\partial \tilde{c})+ (\tilde{b}c \partial \tilde{c})+(\tilde{b}\tilde{c}\partial c)\right) \,.
\ee
This BRST operator squares to
\be 
\frac{Q^2}{\hbar^2} =\frac{1}{12}(c_\text{matter}-52) \oint \mathrm{d}z\, (\partial^3 c c)-\frac{26}{3} h^\vee_\mathfrak{g} \oint  \mathrm{d}z\, (\partial^3 c \tilde{c})-\frac{26}{3} (h_\mathfrak{g}^\vee)^2 \oint  \mathrm{d}z\, (\partial^3 \tilde{c} \tilde{c}).
\ee
Hence the BRST operator is anomalous unless $h^\vee_\mathfrak{g}=0$ and $c_\text{matter}=52$. The critical central charge is the same as in flat space \cite{Mason:2013sva}.
We should note that since $Q^2$ is proportional to $\hbar^2$, the anomaly should be viewed as a quantum correction.

\subsection{Coset manifold}
Finally, we generalize further to coset manifolds, which is the case of interest to describe AdS.
We gauge a subgroup $\text{H} \subset \text{G}_\text{R}$, so that full $\text{G}_\text{L}$ symmetry is preserved, but $\text{G}_\text{R}$ symmetry is broken to the commutant with $\text{H}$.

\paragraph{Classical theory.} We again start by describing the classical action. We introduce a gauge field $A$ for the $\text{H}$ subgroup and the action becomes
\be 
S=\frac{1}{2\pi\hbar} \int \tr(p (g^{-1} \bar{\partial} g-A))+S_\text{current}^{(1)}+S_\text{current}^{(2)}\,.
\ee
The gauge field is a $(0,1)$ form on the worldsheet and hence has no kinetic term.
Under $\text{H}\subset \text{G}_\text{R}$ transformations, the fields transform as
\begin{subequations}
\begin{align}
A &\mapsto hAh^{-1}-\bar{\partial}h h^{-1}\,, \\
g &\mapsto g h^{-1}\,, \\
p &\mapsto h p h^{-1}\,, 
\end{align}
\end{subequations}
which can easily checked to keep the action invariant. The other symmetries go through as before. 

Reparametrization act as
\begin{subequations}
\begin{align}
\delta A&= v \partial A\,, \\
\delta g&= v \partial g \,, \\
\delta p &= \partial (vp)\,.
\end{align}
\end{subequations}
The corresponding conserved current is the energy-momentum tensor that still takes the form $T=\tr(p g^{-1} \partial g)$. The additional gauging does not change the relevant energy-momentum tensor.
The additional ambitwistor symmetry acts exactly as before and hence also $H=\tr(p^2)$ is unchanged.

The various equations of motion read
\begin{subequations}
\begin{align}
\bar{\partial} g-g A&=0\,, \\
\pi(p)&=0\,, \\
\bar{\partial}(g p g^{-1})&=0\,.
\end{align}
\end{subequations}
Here, $\pi$ is the orthogonal projection (orthogonality is defined by the trace) to the subalgebra $\mathfrak{h}$.

\paragraph{Quantum theory.} Finally, we again analyze the quantum theory.  The energy-momentum tensor is the same as for the group
\begin{subequations}
\begin{align}
T&=(J_\text{L}^a j_{\text{L}}^a)-\hbar \, h_\mathfrak{g}^\vee (j_{\text{L}}^a j_{\text{L}}^a)+T_\text{current}\,, \\
H&= (J_\text{L}^a J_\text{L}^a) \,.
\end{align}
\end{subequations}
The gauge algebra gets additionally enhanced by $J^r_\text{R}$. The full gauge algebra reads
\begin{subequations}
\begin{align}
T(z)T(w)&\sim \frac{\hbar^2 c_\text{matter}}{2(z-w)^4}+\frac{2\hbar T(w)}{(z-w)^2}+\frac{\hbar \partial T(w)}{z-w}\ , \\
T(z)H(w)&\sim \frac{2\hbar H(w)}{(z-w)^2}+\frac{\hbar \partial H(w)}{z-w}\ , \\
H(z)H(w)&\sim \frac{4\hbar^2 h_\mathfrak{g}^\vee H(w)}{(z-w)^2}+\frac{2 \hbar^2h_\mathfrak{g}^\vee\partial H(w)}{z-w}\ , \\
T(z) J_\text{R}^r(w) &\sim \frac{\hbar J_\text{R}^r(w)}{(z-w)^2}+\frac{\hbar \partial J_\text{R}^r(w)}{z-w}\ , \\
H(z) J_\text{R}^r(w) &\sim \frac{2\hbar^2 h_\mathfrak{g}^\vee J_\text{R}^r(w)}{(z-w)^2}+\frac{2\hbar^2 h_\mathfrak{g}^\vee \partial J_\text{R}^r(w)}{z-w}\ , \\
J_\text{R}^r(z) J_\text{R}^s(w) &\sim \frac{\hbar\tensor{f}{^{rs}_t} J_\text{R}^t(w)}{z-w}\ .
\end{align}
\end{subequations}
The constraints are therefore first class. We additionally introduce $bc$-ghosts $b_r$ and $c_r$.\footnote{Ghosts with indices refer to the subgroup gauging and ghosts without indices to the ambitwistor gauging, so there should be no confusion.} Since the gauge algebra closes on itself, it is still straightforward to write down a BRST operator. It takes the form
\begin{multline} 
Q=\oint \mathrm{d}z \ \Big(c T+\hbar^{-1}\tilde{c}H+2h^\vee_\mathfrak{g} (\tilde{b}\tilde{c}\partial \tilde{c})+ (\tilde{b}c \partial \tilde{c})+(\tilde{b}\tilde{c}\partial c)\\
+(c_r(J_\text{R}^r+\tfrac{1}{2} J_\text{gh}^r))+(c\partial c_r b_r)+2h_\mathfrak{g}^\vee (\tilde{c}\partial c_r b_r)\Big) \,,
\end{multline}
where
\be 
J_\text{gh}^r=-\tensor{f}{^r_{st}} (c^s b^t)\,.
\ee
The current part of the BRST operator was introduced in \cite{Hlousek:1986ux}.\footnote{This construction is \emph{not} equivalent to the GKO construction of 2d CFT \cite{Goddard:1984vk}, for whose BRST construction an additional $\text{H}$-current needs to be introduced \cite{Karabali:1989dk}.}
This BRST operator squares to
\begin{multline} 
\frac{Q^2}{\hbar^2}=\frac{1}{12}(c_\text{matter}-2\, \text{dim}(\text{H})-52) \oint \mathrm{d}z\, (\partial^3 c c)-\frac{2}{3}(13+\text{dim}(\text{H}) )h^\vee_\mathfrak{g} \oint \mathrm{d}z\, (\partial^3 c \tilde{c})\\
-\frac{2}{3}(13+\text{dim}(\text{H}) )(h^\vee_\mathfrak{g} )^2\oint \mathrm{d}z\, (\partial^3 \tilde{c} \tilde{c})+2h^\vee_\mathfrak{h} \oint \mathrm{d}z\, (\partial c_r c_r)\ . \label{eq:coset Q^2}
\end{multline}
\paragraph{Anomalies.}
We see that the BRST operator is anomalous in most cases of interest. Some of these anomalies can be easily cancelled while others turn out to be more serious for our purpose. The first term is the familiar Weyl anomaly and can be canceled e.g.\ by choosing an appropriate level of the internal current algebra. On the other hand the other three terms are more serious since they impose stringent restrictions on the allowed coset manifolds. 

\paragraph{Ghosts.}
Let us briefly comment about the ghosts in the theory. All the ghost currents are anomalous in the quantum theory. In order for correlation functions to be non-vanishing on a genus $\text{g}$ Riemann surface, we will need the 
\be 
N_c-N_b=N_{\tilde{c}}-N_{\tilde{b}}=3-3\text{g}\,, \qquad N_{c_r}-N_{b_r}=1-\text{g}\,.
\ee
Since we will work on the sphere, we will need to insert as usual 3 $c$-ghosts and $\tilde{c}$-ghosts. We also should insert every $c_r$-ghost once. Since the $c_r$-ghosts have vanishing conformal weight, their one-point function is constant and thus we will ignore this subtlety, but secretly all correlation functions will have an insertion of $\prod_r c_r$.

\section{\label{sec:bosonic-AdS}Bosonic ambitwistor string on AdS}
We now restrict ourselves to AdS. This means that we take
\be 
\text{G}=\text{SO}(d,2)\, , \qquad \text{H}=\text{SO}(d,1)\, .
\ee
In particular, we have $h^\vee_\mathfrak{g}=d$ and $h^\vee_\mathfrak{h}=d-1$.

\subsection{\label{sec:embedding-space}Embedding space formalism}
Before delving into details of ambitwistor string on AdS, let us briefly review the so-called embedding space formalism \cite{Dirac:1936fq,Costa:2011mg, Costa:2011dw}, which allows us to treat the coordinates both in AdS and its boundary in a manifestly covariant manner.

The key idea is to realize AdS$_{d+1}$ as the following hypersurface inside $\mathbb{R}^{2,d}$:
\be
X\cdot X=-1\,.
\ee
Here $X$ is a vector in $\mathbb{R}^{2,d}$, $X^{A}\equiv (X_{-1},X_{0},\ldots, X_{d-1}, X_{d})$ and the inner product is defined by
\be
X\cdot Y\equiv -X_{-1}Y_{-1}-X_{0}Y_{0}+\sum_{j=1}^{d}X_{j}Y_j\,.  
\ee
They are related to standard Poincare coordinates,
\be
ds_{\rm AdS}^2 =\frac{d{\sf z}^2+d{\sf x}^{\mu}d{\sf x}_{\mu}}{{\sf z}^2}\,,
\ee
by the following relation:
\be
X^{A}=\frac{1}{{\sf z}}\left(\frac{1+{\sf z}^2+{\sf x}^{\mu}{\sf x}_{\mu}}{2},\,\,\, {\sf x}^{\mu},\,\,\,\frac{1-{\sf z}^2-{\sf x}^{\mu}{\sf x}_{\mu}}{2}\right)\,.
\ee

Similarly we can realize the boundary of the AdS space-time where the dual CFT lives as a projective null cone inside $\mathbb{R}^{2,d}$:
\be
P\cdot P=0\,,\qquad P^{A}\sim \lambda P^{A}\,,
\ee
for $\lambda \in \mathds{R} \setminus \{0\}$.
To make contact with the standard flat coordinates of the boundary $ds_{\mathbb{R}^{1,d-1}}=d{\sf x}^{\mu}d{\sf x}_{\mu}$, we consider the following parameterization
\be
P^{A}=\left(\frac{1+{\sf x}^{\mu}{\sf x}_{\mu}}{2},\,\,\, {\sf x}^{\mu},\,\,\,\frac{1-{\sf x}^{\mu}{\sf x}_{\mu}}{2}\right)\,.
\ee

The main advantage of using this formalism is that the conformal group SO$(2,d)$ acts linearly on the embedding space coordinates $X^{A}$ and $P^{A}$. Consequently, the generators of the conformal group are given by the following simple expressions
\begin{align}
&D^{a}_{X}\equiv D^{[AB]}_{X}=X^{A}\partial_{X_B}-X^{B}\partial_{X_A}\,,\\
&D^{a}\equiv D^{[AB]}=P^{A}\partial_{P_B}-P^{B}\partial_{P_A}\,. \label{eq:conformal generators}
\end{align}
Here and below we express the indices for the generators, such as $[AB]$ (where the bracket means that the expression is antisymmetric in the two indices), collectively by $a$.
In terms of the embedding space coordinates, the distance between two boundary points can be expressed as
\be
P_{ij}\equiv -2 P_i\cdot P_j =|{\sf x}_i-{\sf x}_j|^2\,,
\ee
while the bulk-to-boundary propagator for an operator with dimension $\Delta$ is given by (up to a constant of proportionality)
\be
K_{\Delta}(X,P)\equiv \frac{1}{(-2 X\cdot P)^{\Delta}}\,.
\ee

\subsection{Vertex operators}
Note the the following anticommutators between the BRST operator and the modes of the $b$-ghosts:
\begin{subequations}
\begin{align} 
\{Q,b_n\}&=\hbar L_n^{\text{tot}}\,, \\
\{Q,\tilde{b}_n\}&=H_n+\hbar\left((c \partial \tilde{b}+2 \partial c \tilde{b})_n-2h_\mathfrak{g}^\vee((b_r \partial c_r)+2(\tilde{b}\partial \tilde{c})+(\partial \tilde{b}\tilde{c}))_n\right)\,, \\
\{Q,b_{r,n}\}&=\hbar\left(J_{\text{R},n}^r+J_\text{gh}^r+(c \partial b_r+\partial c b_r)_n+2d(\tilde{c} \partial b_r+\partial \tilde{c} b_r)_n\right)\,.
\end{align}
\end{subequations}
where $L_n^\text{tot}$ is now the total Virasoro algebra involving all the ghosts. $Q$-closure also requires vertex operators (that sit in the vacuum w.r.t.~the ghosts) to be annihilated by the right-hand side for $n \ge 0$. 

Standard vertex operators take the form
\be 
c\tilde{c}\mathcal{V}^{\alpha\tilde{\alpha}}=c\tilde{c}K^\alpha \tilde{K}^{\tilde{\alpha}}V\,.
\ee
Here, $\alpha$, $\tilde{\alpha}$ are adjoint indices for the two internal current algebras and $K^\alpha$, $\tilde{K}^{\tilde{\alpha}}$ the associated currents.
The above commutation relations imply that $V$ is a primary vertex operator of conformal weight 0. It satisfies
\begin{subequations} \label{eq:vertex operator constraint}
\begin{align}
L_nV&=0 \,,\qquad n\ge 0\,, \\
H_n V&=0 \,,\qquad n> 0\,,\qquad H_0 V=2d\hbar^2  V\,, \\
J_{\text{R},n}^r V&=0 \,,\qquad n\ge 0\,.
\end{align}
\end{subequations}
\paragraph{Inconsistencies and the classical limit.} As it stands the model is inconsistent: the BRST operator does not square to zero and the mass-shell condition of $H$ has an anomalous dimension which will render the path integral inconsistent. We did not find a way to cure these inconsistencies, but we will not be interested in the model per se, but only in the tree level scattering formula it provides. We will not take it too seriously, since we will give an independent proof of our CHY-like formula for AdS. 

We should note that these inconsistencies should perhaps not come as a surprise. Requiring anomaly cancellation gives us the equations of motion of the respective (super)gravity. However, $\text{AdS}_{d+1}$ is not a valid background of the corresponding low-energy theory and thus no consistent (ambitwistor) string theory can be formulated on it. 

In the following we shall take the ambitwistor string as a motivation with the understanding that it does not define a consistent theory. To make sense of it, we are forced to take a classical limit on the worldsheet. We are ignoring anomalous terms in the following and in particular use the non-anomalous mass-shell condition $H_0 V=0$. 

\paragraph{Evaluation of the mass-shell condition.} Let us fix a vector $R \in \mathds{R}^{2,d}$ with $R \cdot R=-1$. Then we can realize $\text{H}=\text{SO}(1,d) \subset \text{G}_\text{R}=\text{SO}(2,d)$ as the stabilizer of $R$. $V$ is a primary of vanishing conformal weight and thus it should be a function of the group-valued field $g$. In order to be invariant under $\text{H}$, it actually only depends on the combination $g R$.
Let us also introduce a vector $P\in \mathds{R}^{2,d}$ and make the following ansatz for the vertex operator
\be 
V(z)\equiv V(P,z)=f(P\tran g R)(z)
\ee
for some function $f$. Since $g$ has regular OPE with itself, there is no normal-ordering issue. This ansatz satisfies all the constraints \eqref{eq:vertex operator constraint} automatically, except for $H_0 V(P,z)=0$. Under $\text{G}_\text{L}$ action, $V(P,z)$ transforms as
\be 
V(P,z) \longmapsto f(P\tran g_\text{L}g R)(z)=V(g_\text{L}^{-1} P,z)\,.
\ee
This implies the OPE
\be 
J_\text{L}^a(z) V(P,z) \sim -\frac{D^a V(P,w)}{z-w}\,,
\ee
where $D^a\equiv D^{[AB]}$ are the conformal generators \eqref{eq:conformal generators} acting on $P$. Up to this point, $P$ is just a formal variable, but it will eventually be interpreted as the boundary coordinate of the vertex operator.
It follows,
\be 
H_0 V(P,z)\propto D^a D^a V(P,z)\,,
\ee
since $H(z)=(J_\text{L}^aJ_\text{L}^a)=(J_\text{R}^aJ_\text{R}^a)$. Denoting $X \equiv g R$, we hence need to solve
\begin{align}
\begin{aligned}
0&= (-P^A \partial_{P_B}P^A \partial_{P_B}+P^A \partial_{P_B}P^B \partial_{P_A})f(P \cdot X) \\
&=-(P \cdot P)(X \cdot X) f''(P \cdot X)+(d+1) P \cdot X f'(P \cdot X) +(P \cdot X)^2 f''(P \cdot X)\,.
\end{aligned}
\end{align}
In order for this to be solvable for any $P$ and $X$, we need to assume that $P \cdot P=0$, i.e.,~$P$ is light-like.\footnote{Since $X\cdot X =R\cdot R=-1$, the only other possibility is $f''(P \cdot X)=0$, however its only solution is a trivial vertex operator equal to the identity.} Then $f(x)$ satisfies the differential equation
\be 
x^2 f''(x)+(d+1)x f'(x)=0\,,
\ee 
and so
\be 
f(x)=A+B x^{-d}\,.
\ee
The constant $A$ leads to the identity vertex operator, which we discard.
We notice that the solutions take the form of a bulk-to-boundary propagator of a massless scalar particle of dimension $d$.

Thus, in the classical limit we are considering, vertex operators take the form
\be 
\mathcal{V}^{\alpha\tilde{\alpha}}=c\tilde{c} K^\alpha \tilde{K}^{\tilde{\alpha}} (-2P\tran g R)^{-d}(z)\,.
\ee
The latter factor is exactly the bulk-to-boundary propagator in AdS (and we chose the normalization factor to agree with it). 

We recognize that $P$ has exactly the same properties as the embedding coordinate $P$ introduced in Section~\ref{sec:embedding-space}. Since the vertex operator turned out to be homogeneous in $P$, it also makes sense to projectivize $P$. We will hence identify $P$ with the coordinate of the dual conformal field theory.
We should mention that worldsheet vertex operators typically take the form of bulk-to-boundary propagators; in the case of $\text{AdS}_3$ this was observed in \cite{deBoer:1998gyt}.
\subsection{Correlation functions}
\paragraph{Localization and scattering equations.}
Let us now discuss the most crucial property of ambitwistor strings. We decompose the Beltrami differential $\tilde{e}$ as follows:
\be 
\tilde{e}=\sum_{i=2}^{n-2}\tilde{e}_i \mu_i\,.
\ee
Here, $\mu_i$ is a basis for the $n-3$ Beltrami differentials on the sphere (the slightly unusual labeling should become clear in the following). Analogously to the flat space situation, the path integral for the $n$-point sphere correlator over the various fields reduces to \cite{Mason:2013sva, Ohmori:2015sha}
\begin{multline} 
\int_{T^*\mathcal{M}_{0,n}} \prod_{i=2}^{n-2}  \mathrm{d}e_i\, \mathrm{d}\tilde{e}_i \ \mathrm{e}^{\sum_i  \tilde{e}_i \int_\Sigma  \mu_i \tr(p^2)} \\
\times\left\langle c\tilde{c}\mathcal{V}_{1}^{\alpha_1\tilde{\alpha}_1}(0)\prod_{i=2}^{n-2} \mathcal{V}_i^{\alpha_i\tilde{\alpha}_i}(z_i)\; c\tilde{c}\mathcal{V}_{n-1}^{\alpha_{n-1}\tilde{\alpha}_{n-1}}(1)\; c\tilde{c}\mathcal{V}_n^{\alpha_n\tilde{\alpha}_n}(\infty) \right\rangle_\Sigma \,,
\end{multline}
where the correlation function in the integrand is understood to be taken in the worldsheet CFT.
Here, the Beltrami differentials $e$ parametrize deformations in $\mathcal{M}_{0,n}$ and $\tilde{e}$ the cotangent space. Due to our choice of gauge of the conformal Killing vectors, the $c$ and $\tilde{c}$ correlators are trivial. Let us make a standard choice for the basis of Beltrami differentials:
\be 
\int_\Sigma \mu_i \tr(p^2)=\mathop{\text{Res}}_{z=z_i} \tr(p^2)\,.
\ee
Thus it remains to evaluate this residue when acting on the correlator. Recalling that $\tr(p^2)=\tr(J_\text{L} J_\text{L})$, this is precisely computed by the Knizhnik-Zamolodchikov equation in the quantum theory \cite{Knizhnik:1984nr}. We hence have\footnote{We note that this would lead to inconsistencies if we are working at finite $\hbar$, since the anomalous mass-shell condition of $H(z)$ means that $\tr(p^2)$ has a double pole and hence cannot be contracted with a Beltrami differential.}
\be\label{eq:scattering-equations}
\mathop{\text{Res}}_{z=z_i} \tr(p^2)=-\sum_{\substack{j=1\\ j \ne i}}^{n}\frac{D_i \cdot D_j}{z_i-z_j}\equiv -\frac{1}{2} E_i
\ee
as an operator acting on the correlation function. 

We can then integrate out $\tilde{e}_i$, which leads to the factor
\be 
\bigwedge_{i=2}^{n-2}\bar{\delta}(E_i)\,.
\ee
This is an operator valued $\delta$-function. We can define
\be 
\bar{\delta}(E_i)\equiv -\frac{1}{2\pi i} \bar{\partial} \frac{1}{E_i} \,,
\ee
which defines a $(0,1)$-form. 
Thus, the ambitwistor string correlator becomes
\be 
\int_{\mathcal{M}_{0,n}} \bigwedge_{i=2}^{n-2} \mathrm{d} z_i\, \bar{\delta}(E_i)\left\langle \mathcal{V}_{1}^{\alpha_1\tilde{\alpha}_1}(0)\prod_{i=2}^{n-2} \mathcal{V}_i^{\alpha_i\tilde{\alpha}_i}(z_i)\; \mathcal{V}_{n-1}^{\alpha_{n-1}\tilde{\alpha}_{n-1}}(1)\mathcal{V}_n^{\alpha_n\tilde{\alpha}_n}(\infty) \right\rangle\,.
\ee
The integrand is an $(n{-}3,n{-}3)$-form as is appropriate for the integration.

\paragraph{Rewriting as contour integral.} For later purposes, it is useful to rewrite the result as a purely holomorphic contour integral. Using the definition of $\bar{\delta}$, we can write
\begin{align} 
\int_{\mathcal{M}_{0,n}} \bigwedge_{i=2}^{n-2} \mathrm{d} z_i\, \bar{\delta}(E_i)\, \la \cdots\ra
&=-\frac{1}{2\pi i} \int_{\mathcal{M}_{0,n}}   \bar{\partial} \left(\frac{1}{E_2}\bigwedge_{i=3}^{n-2}\mathrm{d} z_i\, \bar{\delta}(E_i)\, \la \cdots \ra \right)\\
&=-\frac{1}{2\pi i} \oint_{\{|E_2|=\varepsilon\} \cap \mathcal{M}_{0,n}}  \frac{1}{E_2}\bigwedge_{i=3}^{n-2}  \d z_i\,  \bar{\delta}(E_i)\, \la \cdots \ra,
\end{align}
which using integration by parts encloses $E_2 = 0$, which is the only place where the integrand fails to be exact. We abbreviated the holomorphic CFT correlator to $\la \cdots \ra$.

Let us explain what exactly we mean by $E_2=0$, since $E_2$ is operator-valued. The operator $E_2^{-1}$ acts on the CFT correlation function $\langle \cdots \rangle$ and produces some analytic function in $z_2,\dots,z_{n-2}$, which has  additional poles beyond the poles already present in the original correlators. The set $E_2=0$ denotes these additional poles. Since $E_2$ is operator-valued, we do not have a good understanding how many poles $E_2=0$ comprises in general, but this knowledge will not be needed in the following. We will repeatedly write such expressions in the following and they are always meant to be understood in this way.

Continuing recursively in the same manner, we arrive at
\be 
\int_{\mathcal{M}_{0,n}} \ \bigwedge_{i=2}^{n-2} \mathrm{d} z_i\, \bar{\delta}(E_i)\, \la \cdots \ra
=(-1)^{n-3}  \oint_\Gamma  \bigwedge_{i=2}^{n-2} \frac{\mathrm{d} z_i}{2\pi i E_i}\, \la \cdots\ra\,,
\ee
where $\Gamma = \cap_{i=2}^{n-2} \{ |E_i| = \varepsilon \}$. A more rigorous derivation using Morse theory of this formula was given in \cite{Ohmori:2015sha}. One can directly copy the argument to our setting. Essentially, we see that the structure of the moduli space integral is unchanged, except that $\tr(p^2)$ is operator-valued.
\paragraph{CFT correlators and the contact diagram.}
It remains to evaluate the CFT part of the correlation function. It clearly factorizes into the current part and the coset part.
Removing as usual double contractions \cite{Casali:2015vta}, the current correlators become a product of Parke-Taylor factors $\text{PT} \times \tilde{\text{PT}}$, where
\be 
\text{PT}=\sum_{\sigma \in S_n / \mathds{Z}_n} \frac{\text{Tr}\,(T^{\alpha_{\sigma(1)}} \cdots T^{\alpha_{\sigma(n)}})}{z_{\sigma(1)\sigma(2)} \cdots z_{\sigma(n-1)\sigma(n)} z_{\sigma(n)\sigma(1)}}\,,
\ee
where $z_{ij} = z_i - z_j$.
Here, $T^\alpha$ are the generators of the internal group in the adjoint representation. 
Thus, the remaining correlator becomes
\be 
\langle V(P_1,z_1) V(P_2,z_2) \cdots V(P_n,z_n)\rangle\,,
\ee
where $V(P,z)=(-2 P\tran g R)^{-d}(z)$, as discussed above. This correlator can again be evaluated by a path-integral computation. The equations of motion imply that $g$ is holomorphic. Moreover, since its OPE with $g$ vanishes, it has no singularities close to the insertion points. By compactness of the Riemann sphere, $g$ hence has to be constant in this correlator. Thus, the path integral reduces to an ordinary integral over the group $\text{SO}(d,2)$ (with the Haar measure as measure). 
By construction, the integral is invariant under right multiplication of $g$ by the subgroup $\mathrm{SO}(1,d)$ and thus the integral reduces further to the coset manifold $\text{AdS}_{d+1}$. The measure is the induced one, which is the unique (up to a normalization constant) left-invariant measure in $\text{AdS}_{d+1}$. Writing $X =g R$ as above, the correlation function hence equals
\be 
\contact(P_1,\dots,P_n)=\langle V(P_1,z_1) V(P_2,z_2) \cdots V(P_n,z_n)\rangle=\int_{\text{AdS}_{d+1}} \mathrm{d}X \ \prod_{i=1}^n \frac{1}{(-2X \cdot P_i)^d}\,.
\ee
This is nothing else than the scalar contact diagram in $\text{AdS}_{d+1}$ \cite{Freedman:1998tz}. Note in particular that this is independent of the worldsheet insertions.

\paragraph{Summary.} Let us summarize the result and reinstate $\text{SL}(2,\mathds{C})$ invariance. The ambitwistor string correlator of the basic vertex operators that we have analyzed gives
\be\label{eq:taylorAdS}
m_\text{AdS}=\sum_{\sigma,\, \tau \in S_n/\mathds{Z}_n} m_\text{AdS}(\sigma|\tau)\text{Tr}\,(T^{\alpha_{\sigma(1)}} \cdots T^{\alpha_{\sigma(n)}})\text{Tr}\,(T^{\tilde{\alpha}_{\tau(1)}} \cdots T^{\tilde{\alpha}_{\tau(n)}})\,,
\ee
with
\begin{align}
m_\text{AdS}(\sigma|\tau)&=\int
\frac{(z_{1,n-1}z_{n-1,n}z_{1n})^2}{\prod_{i=1}^{n} z_{\sigma(i)\sigma(i+1)} \prod_{j=1}^{n} z_{\tau(j)\tau(j+1)}}
\bigwedge_{i=2}^{n-2} \d z_i\, \bar{\delta}(E_i)\,  \contact\,, \\
 &= (-1)^{n-3} \!\! \oint_{\Gamma} \frac{1}{\prod_{i=1}^{n} z_{\sigma(i)\sigma(i+1)} \prod_{j=1}^{n} z_{\tau(j)\tau(j+1)}}  \frac{1}{J} \bigwedge_{i=1}^{n} \frac{\d z_i}{2\pi i \, E_i} \,\contact\,,\label{m-AdS}
\end{align}
where $\sigma(n{+}1)=\sigma(1)$ and $\tau(n{+}1) = \tau(1)$.
Let us explain all factors in \eqref{m-AdS} in turn. The integral is defined on the moduli space of genus zero curves with $n$ marked points (punctures) with coordinates $z_i$. By convention we fix the positions of three points, $(z_1, z_{n-1}, z_n)$, using the action of $\SL(2,\C)$. In doing so we obtain a Jacobian factor:
\be 
J = \frac{1}{(2\pi i)^3 E_1 E_{n-1} E_{n}} \frac{\d z_1 \wedge \d z_{n-1} \wedge \d z_n}{(z_{1,n-1}z_{n-1,n}z_{1n})^2}\,,
\ee
which in particular makes the integrand a top holomorphic form on $\mathcal{M}_{0,n}$. As before, we have
\be
E_i = \sum_{\substack{j=1\\ j \neq i}}^{n} \frac{2D_i \cdot D_j}{z_i - z_j}
\ee
and the contour of integration $\Gamma = \cap_{i=2}^{n-2} \{ |E_i| = \varepsilon \}$, whose orientation is induced from $\mathcal{M}_{0,n}$,
imposes the scattering equations. (In particular we will make a repeated use of the fact that with this assignment of orientation the individual contours anticommute, $\Gamma_1 \cap \Gamma_2 = - \Gamma_2 \cap \Gamma_1$.) These factors act on $\contact\equiv \contact(P_1,\dots,P_n)$, which denotes the contact scalar diagram. Recall that this term is $z$-independent.

Let us again emphasize that this formula should be taken with a grain of salt, since the ambitwistor suffers from anomalies. We take it as a motivation and will check in the next section that it is indeed self-consistent and is equivalent to Witten diagrams on $\text{AdS}_{d+1}$.

\paragraph{Relation to the formula of Roehrig-Skinner.} Let us compare this formula with the formula derived in \cite{Roehrig:2019zqb}. Their formula looks visually identical after restricting to $\text{AdS}_3 \times \text{S}^3$, with one important difference. In the case of $\text{AdS}_3$, the conformal group factorizes, $\mathrm{SO}(2,2) \cong \mathrm{SL}(2,\mathds{R}) \times \mathrm{SL}(2,\mathds{R})$ (up to global identifications) and correspondingly we can decompose the conformal generators in $D^a_{+}$ and $D^a_-$. Explicitly,
\be 
D^{AB}_\pm=\frac{1}{2} \left(D^{AB}\pm \frac{1}{2}\varepsilon^{ABCD} D_{CD}\right)\ ,
\ee
where $\varepsilon_{ABCD}$ is the Levi-Civita symbol. The formula of Roehrig and Skinner is identical to ours, except that $D_+^a$ appears in the scattering equation instead of $D^a$. The difference in the formulae comes from a different treatment of $\text{AdS}_3$. We treated it as a coset, whereas it was considered as a group manifold in \cite{Roehrig:2019zqb}.
We now argue that these are equivalent formulae. The basic reason is that the contact diagram contains only scalars in the partial wave decomposition, see e.g.~\cite{Hijano:2015zsa}. For this reason, $D_{+,i} \cdot D_{+,j}$ when acting on the contact diagram equals $D_{-,i} \cdot D_{-,j}$. Thus, we have
\be 
E_i = \sum_{\substack{j=1\\ j \neq i}}^{n} \frac{2D_i \cdot D_j}{z_i - z_j}=E_{+,i} = \sum_{\substack{j=1\\ j \neq i}}^{n} \frac{4D_{+,i} \cdot D_{+,j}}{z_i - z_j}\ ,
\ee
when acting on the contact diagram.

\section{\label{sec:correlators}Computation of the correlators}


In the previous section we found that the ambitwistor formulation gives rise to a CHY formula for the tree-level color-ordered bi-adjoint scalar correlators \eqref{m-AdS}.
Before evaluating the formula \eqref{m-AdS}, let us recall a few facts about the contractions $D_i {\cdot} D_j$.


\subsection{Properties of conformal generators}

Recall from Section~\ref{sec:embedding-space} that we use conformal generators in the scalar representation:
\be
D^{a}_{i} = D^{[AB]}_{i} = P^A_i \partial_{P_i^B}-P^B_i \partial_{P_i^A}\,,
\ee
where $A,B = -1,0,1,\ldots,d$ are the embedding space indices and $i,j=1,2,\ldots,n$ label each external particle. Indices can be contracted according to
\be
D_i \cdot D_j = \eta_{AC} \eta_{BD} D_i^{AB} D_j^{CD}\,.
\ee
In particular, $D_i \cdot D_i = 0$ for any $i$. Trivially $[D_i, D_j] = 0$ for any $i \neq j$.  We will make use of the following notation for multi-particle contractions (we already used the fact that $D_i \cdot D_i=0$ when acting on the contact diagram)
\be
D_{i_1 i_2\ldots i_m}^2 = (D_{i_1} {+} D_{i_2} + \dots + D_{i_m})^2 = \sum_{1 \leq j<k \leq m} 2D_{i_j} \cdot D_{i_k}\,.
\ee
The ordering of individual labels $\{ i_1, i_2,\ldots, i_m\}$ is immaterial.
It is straightforward to see that $D_S^2$ and $D_{T}^2$ commute if and only if the two sets of particles $S$ and $T$ are either disjoint or one is contained within the other,
\be\label{DD-commutator}
[D_S^2, D_T^2] = 0\qquad \text{iff}\qquad S \cap T = \varnothing\quad\text{or}\quad S \subseteq T \quad\text{or}\quad T \subseteq S\,.
\ee

Correlation functions, such as \eqref{m-AdS} above, satisfy the Ward identity:
\be
\sum_{i=1}^{n} D_{i}^{AB} \la \cdots \ra = 0\,.
\ee
In order to see this from the scattering equations formula \eqref{m-AdS}, we need to check that the operator $\sum_i D_i^a$ commutes through the scattering equations (since it already annihilates the contact term $\contact$). It is enough to check that 
\be
\left[ \sum_{i=1}^{n} D_i^a,\, D_j \cdot D_k \right] = [ D_j^a {+} D_k^a,\, D_j \cdot D_k] = \frac{1}{2} [ D_j^a {+} D_k^a,\, (D_j {+} D_k)^2] = 0\,.
\ee
We should mention one regularity assumption that we are making in the following. We take it for granted that the differential operators under consideration, such as $D_i \cdot D_j$ or the scattering equations $E_i$ have an eigenbasis when acting on a suitable function space in the variables $P_1,\dots,P_n$. In the case of $D_i \cdot D_j$, the eigenfunctions are the (higher-point) conformal blocks, but in the case of $E_i$ they are some object that interpolate in between the different channels of the conformal block expansion. This is explored further in Section~\ref{sec:eigenfunctions}, where we analyze these generalized conformal blocks in the simple case of a four-point function. For two commuting differential operators $\mathscr{D}_1$ and $\mathscr{D}_2$, we can hence find a simultaneous eigenbasis. In this joint eigenbasis, it is easy to define $f(\mathscr{D}_1)$ and $g(\mathscr{D}_2)$ for any function (or even a distribution). We hence always assume the general implication
\be 
[\mathscr{D}_1,\mathscr{D}_2]=0 \quad \Longrightarrow \quad [f(\mathscr{D}_1),g(\mathscr{D}_2)]=0\,.
\ee
With this, it follows that $\sum_i D_i^a$ commutes with $\bar{\delta}(E_j)$, which shows that \eqref{m-AdS} satisfies the Ward identity. 

Finally, let us show that different $E_i$'s commute. We start with
\be
[E_i, E_j] = \left[ \sum_{k \neq i} \frac{2D_i \cdot D_k}{z_i - z_k},\; \sum_{l \neq j} \frac{2D_j \cdot D_l}{z_j - z_l} \right]
\ee
for any pair $i,j$.
The only non-vanishing terms on the right-hand side are two corresponding to the diagonal terms $k=l$, as well as $k=j$ in the first sum and $l=i$ in the second. This leaves us with
\be
\tfrac{1}{4}[E_i, E_j] = \sum_{k \neq i,j} \left[ \frac{D_i \cdot D_k}{z_i - z_k},\; \frac{D_j \cdot D_k}{z_j - z_k} \right] + \sum_{l \neq i,j} \left[ \frac{D_i \cdot D_j}{z_i - z_j},\; \frac{D_j \cdot D_l}{z_j - z_l} \right] + \sum_{k \neq i,j} \left[ \frac{D_i \cdot D_k}{z_i - z_k},\; \frac{D_j \cdot D_i}{z_j - z_i} \right]\,.
\ee
Next we relate the three commutators by using structure constants $f_{abc}$ and their antisymmetry properties,
\begin{subequations} 
\begin{align}
[D_i \cdot D_k, \, D_j \cdot D_k] &= f_{abc}\, D_i^{a} D_j^{b} D_k^{c},\\
[D_i \cdot D_j, \, D_j \cdot D_l] &= f_{acb}\, D_i^{a} D_j^{c} D_l^{b} = - [D_i \cdot D_l, \, D_j \cdot D_l]\,,\\
[D_i \cdot D_k, \, D_j \cdot D_i] &= f_{cba}\, D_i^{c} D_j^{b} D_k^{a} = -[D_i \cdot D_k, \, D_j \cdot D_k]\,.
\end{align}
\end{subequations} 
Relabeling $l \to k$ in the second sum yields
\be
\tfrac{1}{4}[E_i, E_j] = \!\!\sum_{k \neq i,j}\! \left( \frac{1}{(z_i {-} z_k)(z_j {-} z_k)} {-} \frac{1}{(z_i {-} z_j)(z_j {-} z_k)} {-} \frac{1}{(z_i {-} z_k)(z_j {-} z_i)}   \right) [D_i \cdot D_k, \, D_j \cdot D_k] = 0\,,
\ee
which vanishes for each term in the sum by a partial fraction identity. This shows that in \eqref{m-AdS} we do not have to worry about the order of $E_i$ and that different choices of $\text{SL}(2,\C)$ fixing are equivalent to each other. We note that the condition $[E_i, E_j] =0$ is equivalent to imposing integrability for the Knizhnik-Zamolodchikov connection that we effectively use in the $\alpha' \to \infty$ limit.

\subsection{Computation with residue theorems}

Let us consider the problem of evaluating the formula \eqref{m-AdS}. We will show that
\be\label{m-AdS-result}
m_{\text{AdS}}(\sigma | \tau) \;= (-1)^{w(\sigma|\tau)+1}\!\!\!\! \sum_{T \in T(\sigma) \cap T(\tau)} \prod_{e \in T} \frac{1}{D_e^2} \;\contact(P_1, P_2,\ldots, P_n)\,,
\ee
where $T(\sigma)$ and $T(\tau)$ are sets of trivalent trees planar with respect to the permutation $\sigma$ and $\tau$ respectively. The sum goes over all such trees $T$ that are simultaneously planar with respect to both permutations. The product involves exactly $n{-}3$ propagators for each internal (bulk-to-bulk) edge $e \in T$ with $D_e$ equal to the sum of $D_i$'s flowing into that edge. The overall sign is determined by the relative winding number of the two permutations $w(\sigma|\tau)$.\footnote{%
It is defined as follows. Consider $n$ points on a circle and label them in a clockwise direction according to the permutation $\sigma$. Then consider a path starting at the label $\tau(1)$ going to $\tau(2)$ in a clockwise direction, then to $\tau(3)$, etc., until the last step where $\tau(n)$ connects back to $\tau(1)$. The number of windings performed by this path defines $w(\sigma | \tau)$. In particular, when $\sigma=\tau$ we have $w(\sigma | \sigma)=1$. In flat space one can show that this definition determines the correct sign for each bi-color ordering agreeing with Feynman diagrams.
}
The whole sum acts on the scalar contact diagram $\contact$. Notice that for every tree the inverse propagators $D_e^2$ precisely satisfy the non-overlapping criterion \eqref{DD-commutator}, which means that the order in which they appear does not matter.

In the special case when $\sigma = \tau$, the above expression simplifies to a sum over planar trees (we also have $w(\sigma|\sigma)=1$). We will focus on this case first, by considering the diagonal entries $m_{\text{AdS}}(\I_n| \I_n)$, where without loss of generality we consider both permutations to be the identity $\I_n = (12 \cdots n)$. The proof of the more general case with $\sigma \neq \tau$ is essentially identical and we will return back to it after spelling out all details for the diagonal case.

The strategy in proving \eqref{m-AdS-result} will be to rewrite the integral localizing on solutions of scattering equations to that localizing on boundaries of $\mathcal{M}_{0,n}$ corresponding to the Riemann surface degenerating into trivalent diagrams. While in flat space case there are multiple ways of arriving at such a result, see, e.g., \cite{Dolan:2013isa,Cachazo:2015nwa,Arkani-Hamed:2017mur,Mizera:2019gea}, here we cannot use them reliably for operator-valued scattering equations. In other words, we do not have a reliable way of solving these constraints explicitly, or even predicting how many solutions they might have. Therefore, closely following \cite{Dolan:2013isa}, we will show the required results using purely contour deformation arguments, which do not rely on any knowledge of solutions of scattering equations.

Before proceeding to the general proof it will be instructive to work out the $n=4,5$ correlation functions first. 

\subsubsection{Four-point correlator}

Fixing $(z_1,z_3,z_4)$ leaves us with $z_2$ as the only leftover variable. We have
\be
m_{\text{AdS}}(\I_4 | \I_4) = -\oint_{\Gamma_2} \frac{(z_{13}z_{34}z_{14})^2}{(z_{12} z_{23} z_{34} z_{41})^2} \frac{\d z_2}{2\pi i E_2}\, \contact\,.
\ee
In order to make the pole structure more manifest we turn $E_2$ into a polynomial $\hat{E}_2$ by
\be
\hat{E}_2 = z_{21} z_{23} z_{24} E_2 = z_{23}z_{24} D_{12}^2 + z_{21} z_{24} D_{23}^2 + z_{21}z_{23} D_{24}^2\,,
\ee
which gives
\be
m_{\text{AdS}}(\I_4 | \I_4) = \oint_{\Gamma_2} \frac{z_{13}^2 z_{24}}{z_{12} z_{23}} \frac{\d z_2}{2\pi i \hat{E}_2} \,\contact\,.
\ee
Therefore the only poles that survive are located at $\hat{E}_2 = 0$, $z_2 = z_1$, and $z_2 = z_3$. The contour above is taken around the first class of poles, $\Gamma_2 = \{ |\hat{E}_2| = \varepsilon\}$. Since in general $\SL(2,\C)$ fixing there cannot be poles at infinity, we deforming the contour we localize on residues around $z_2 = z_1$ and $z_2 = z_3$. These are precisely the places in the moduli space corresponding to the worlsheet degenerating into $s$- and $t$-channel diagrams. In equations, we find
\be
m_{\text{AdS}}(\I_4 | \I_4) = -\left( \oint_{|z_2 - z_1|=\varepsilon} +  \oint_{|z_2 - z_3|=\varepsilon}  \right) \frac{z_{13}^2 z_{24}}{z_{12} z_{23}} \frac{\d z_2}{2\pi i \hat{E}_2}\,\contact  =  \left(\frac{1}{D^2_{12}} + \frac{1}{D^2_{23}}\right) \contact\,,
\ee
as expected. Note that we did not have to have any concrete knowledge of the positions, or even the number, of the solutions to the scattering equations.

The absence of $u$-channel poles is a consequence of the fact that the integrand did not have any pole in $z_2 = z_4$. More generally, the half-integrands need to have a combined double pole in a specific degeneration for the correlators to develop the corresponding $D_e^2$ singularity.

\subsubsection{Five-point correlator}

In the standard gauge fixing we arrive at the following representation,
\be
m_{\text{AdS}}(\I_5 | \I_5) = \oint_{\Gamma_2 \cap \Gamma_3} \frac{(z_{14}z_{45}z_{15})^2}{(z_{12} z_{23} z_{34} z_{45} z_{51})^2} \frac{\d z_2 \wedge \d z_3}{(2\pi i)^2 E_2 E_3}\,\contact\,.
\ee
As before, we make the pole structure manifest by defining
\be\label{E-hats}
\hat{E}_2 = z_{21} z_{23} z_{24} z_{25} E_2, \qquad \hat{E}_3 = z_{31} z_{32} z_{34} z_{35} E_3\,,
\ee
which gives
\be\label{mAdS-5}
m_{\text{AdS}}(\I_5 | \I_5) = \oint_{\Gamma_2 \cap \Gamma_3} \frac{z_{14}^2 z_{24} z_{25} z_{31} z_{35}}{z_{12}  z_{34}} \frac{\d z_2 \wedge \d z_3}{(2\pi i)^2 \hat{E}_2 \hat{E}_3} \,\contact\, .
\ee
Hence the only poles are those corresponding to $z_2 = z_1$, $z_3 = z_4$, on top of those coming from the scattering equations. In these variables the integration contour is simply
\be
\Gamma_2 \cap \Gamma_3 = \{ |\hat{E}_2| = \varepsilon\} \cap \{ |\hat{E}_3| = \varepsilon\}\, .
\ee
Let us also define contours associated to the other singularities, namely
\be
\gamma_{21} = \{ |z_2 - z_1| = \varepsilon \}, \qquad \gamma_{34} = \{ |z_3 - z_4| = \varepsilon\}\, .
\ee
We can now use a higher-dimensional residue theorem, the \emph{global residue theorem} (see \cite[Chapter~5]{griffiths2014principles}), to relate integrals over these contours to each other. In the present application it says that the integral of the same top form over
\be
(\Gamma_2 + \gamma_{34}) \cap (\Gamma_3 + \gamma_{21}) = \Gamma_2 \cap \Gamma_3 + \Gamma_2 \cap \gamma_{21} + \gamma_{34} \cap \Gamma_3 - \gamma_{21} \cap \gamma_{34}
\ee
vanishes, since the pole divisor of our integrand is contained within $\Gamma_2 \cup \Gamma_3 \cup \gamma_{21} \cup \gamma_{34}$. Using this fact we can rewrite the correlator \eqref{mAdS-5} as
\be\label{mAdS-3-residues}
m_{\text{AdS}}(\I_5 | \I_5) = \left( - \oint_{\Gamma_2 \cap \gamma_{21}} -  \oint_{\gamma_{34} \cap \Gamma_{3}} +  \oint_{\gamma_{21} \cap \gamma_{34}} \right) \frac{z_{14}^2 z_{24} z_{25} z_{31} z_{35}}{z_{12}  z_{34}} \frac{\d z_2 \wedge \d z_3}{(2\pi i)^2 \hat{E}_2 \hat{E}_3} \,\contact\, .
\ee
Let us compute these residue integrals in turn. The final one is straightforward to evaluate and gives
\be
\oint_{\gamma_{21} \cap \gamma_{34}} \frac{z_{14}^2 z_{24} z_{25} z_{31} z_{35}}{z_{12}  z_{34}} \frac{\d z_2 \wedge \d z_3}{(2\pi i)^2 \hat{E}_2 \hat{E}_3}\,\contact = \frac{1}{D_{12}^2 D_{34}^2}\,\contact\, ,
\ee
which is the expected answer coming from this worldsheet degeneration.

The evaluation of the first two integrals is a bit more subtle. In order to see this let us focus on the the case $\Gamma_2 \cap \gamma_{21}$. The first part of the contour imposes that
\be
\hat{E}_2 = z_{21} \left( \cdots \right)  + z_{23} z_{24} z_{25} D_{12}^2 = 0\, .
\ee
However, on the additional constraint surface $\gamma_{21}$, for this to be true the combination $z_{23}=z_2 - z_3$ needs to vanish as well (note that $z_{24}$ and $z_{25}$ cannot vanish near $\gamma_{21}$). In other words, as the $z_2$ puncture approaches $z_1$ at some infinitesimal rate $\delta$, the puncture $z_3$ is also forced to $z_2$ at the same rate, so that all three punctures $(z_1, z_2, z_3)$ coalesce together as $\delta \to 0$. We can parameterize this degeneration explicitly by setting
\be\label{cov}
z_i = z_1 + \delta x_i \qquad\text{for}\qquad i=1,2,3
\ee
with $x_1=0$.
Expanding in small $\delta$, the two equations \eqref{E-hats} become
\begin{align}
\hat{E}_i = \delta z_{14} z_{15} \hat{F}_i  + \mathcal{O}(\delta^2) \qquad \text{with}\qquad \begin{array}{l}
\hat{F}_2 = x_{23} D_{12}^2 + x_{21} D_{23}^2\, , \\
\hat{F}_3 = x_{32} D_{13}^2 + x_{31} D_{23}^2\, .
\end{array}
\end{align}
In the new variables $(x_2,\delta)$ we have $\d z_2 \wedge \d z_3 = \delta x_{31}\, \d x_2 \wedge \d\delta$ and the contour becomes $\{ |\hat{F}_2 | = \varepsilon\} \cap \{ |\delta| = \varepsilon\} $. Therefore the first contribution to \eqref{mAdS-3-residues} becomes:
\begin{align}
-\oint_{|\hat{F}_2| = \varepsilon} \oint_{|\delta| = \varepsilon} \!\! \left( \frac{x_{31}^2}{\delta\, x_{12}} + \mathcal{O}(\delta^0)  \right)\frac{\d x_2 \wedge \d\delta}{(2\pi i)^2 \hat{F}_2 \hat{F}_3}\,\contact = -\oint_{|\hat{F}_2| = \varepsilon} \frac{x_{31}^2}{x_{12}} \frac{\d x_2}{2\pi i\, \hat{F}_2 \hat{F}_3} \contact \, .
\end{align}
At this stage we consider the difference
\be
\hat{F}_2 - \hat{F}_3 = x_{23}(D_{12}^2 + D_{13}^2) + (x_{21} - x_{31}) D_{23}^2 = x_{23} D_{45}^2\, ,
\ee
where in the last transition we used the Ward identity. Therefore we conclude that on the support of the constraint $\hat{F}_2 = 0$ we have $\hat{F}_3 = x_{23} D_{45}^2$. Hence the above integral simplifies to
\begin{align}
-\oint_{|\hat{F}_2| = \varepsilon} \frac{x_{31}^2}{x_{12} x_{23}} \frac{\d x_2}{2\pi i\, \hat{F}_2} \frac{1}{D_{45}^2} \contact &= \left( \oint_{|x_2 - x_1| = \varepsilon} + \oint_{|x_2 - x_3| = \varepsilon}  \right) \frac{x_{31}^2}{x_{12} x_{23}} \frac{\d x_2}{2\pi i\, \hat{F}_2} \frac{1}{D_{45}^2} \contact \\
&=  \!\left( \frac{1}{D_{12}^2} + \frac{1}{D_{23}^2} \right) \frac{1}{D_{45}^2}\contact\, ,
\end{align}
where in the first equality we used a residue theorem to deform the contour to enclose the poles at $x_2 = x_1$ and $x_2 = x_3$ (the one at infinity is absent), similarly to the $n=4$ case.
Let us remark that while the above multi-step procedure might seem complicated, it is an unavoidable property of $\mathcal{M}_{0,n}$ itself: the manual change of variables we used in \eqref{cov} implements a blow-up procedure needed for compactifying the moduli space \cite{PMIHES_1969__36__75_0}.

Evaluation of the remaining contribution from $\gamma_{34} \cap \Gamma_3$ proceeds entirely analogously. As a shortcut we can exploit symmetry of the problem under $(12345) \to (43215)$ to write immediately:
\be
-  \oint_{\gamma_{34} \cap \Gamma_{3}}  \frac{z_{14}^2 z_{24} z_{25} z_{31} z_{35}}{z_{12}  z_{34}} \frac{\d z_2 \wedge \d z_3}{(2\pi i)^2 \hat{E}_2 \hat{E}_3}\,\contact = \frac{1}{D_{51}^2} \left( \frac{1}{D_{23}^2} + \frac{1}{D_{34}^2} \right)\contact\, .
\ee
Summing all contributions in \eqref{mAdS-3-residues} we found
\be
m_{\text{AdS}}(\I_5 | \I_5) = \left(\frac{1}{D_{12}^2 D_{34}^2} +  \frac{1}{D_{23}^2 D_{45}^2} +  \frac{1}{D_{34}^2 D_{51}^2} +  \frac{1}{D_{45}^2 D_{12}^2} +  \frac{1}{D_{51}^2 D_{23}^2}\right) \contact\, ,  
\ee
which according to the expectation is a sum over all planar cubic diagrams.

\subsubsection{Arbitrary-multiplicity correlators}

Proof in the general case uses essentially the same steps and is mostly an exercise in bookkeeping. Following \cite{Dolan:2013isa} we will achieve this by showing that \eqref{m-AdS} obeys the Berends-Giele (off-shell) recursion relation:
\begin{align}
\hat{m}_{\text{AdS}}(\I_n | \I_n) &= \frac{1}{D_{n-1,n}^2} \hat{m}_{\text{AdS}}(12\cdots n{-}2, (n{-}1,n) \,|\, 12\cdots n{-}2, (n{-}1,n) ) \nn\\
&+ \frac{1}{D_{n1}^2} \hat{m}_{\text{AdS}}(2\cdots n{-}1,(n1) \,|\, 2\cdots n{-}1,(n1) ) \nn\\
&+ \sum_{k=2}^{n-3} \frac{1}{D_{12\dots k}^2\, D_{k+1, k+2, \dots, n-1}^2} \hat{m}_{\text{AdS}}( 12\cdots k (k{+}1,\cdots n)\,|\, 12\cdots k (k{+}1,\cdots n) )\nn\\
&\  \times \hat{m}_{\text{AdS}}(k{+}1,k{+}2,\cdots,n{-}1(n1\cdots k)\,|\, k{+}1,k{+}2,\cdots,n{-}1(n1\cdots k))\, .\label{recursion-relation}
\end{align}
\begin{figure}
	\centering
	\includegraphics{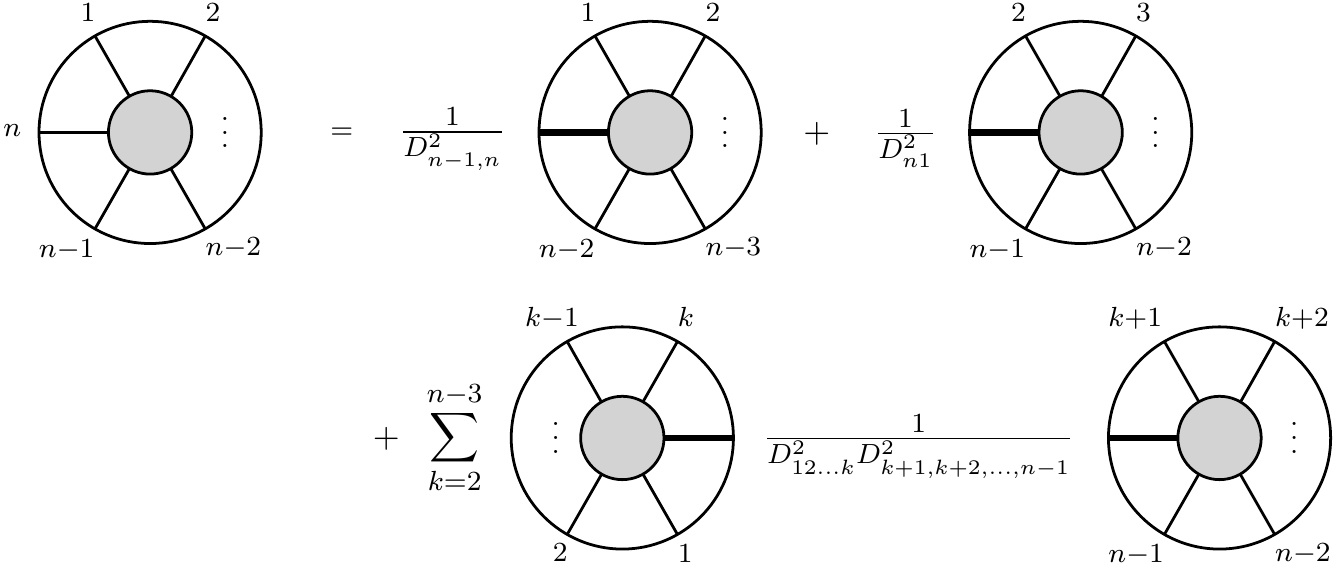}
	\caption{\label{fig:recursion}Diagrammatic depiction of the recursion relation from \eqref{recursion-relation}. The new external states (bold) have conformal generators determined by the Ward identities $\sum_i D_i = 0$ for each term on the right-hand side individually.}
\end{figure}
Here terms in the parenthesis denote a single particle with `momentum' equal to the sum of all labels within the parenthesis, e.g., the first contribution is a correlator with $n{-}1$ external states, the last of which has the conformal generator $D_{n-1}{+}D_n$.
The recursion might be easier to understand from its diagrammatic representation in Figure~\ref{fig:recursion}.
The reason why the $n$-th state appears special in the recursion is ultimately a consequence of our $\SL(2,\C)$ fixing. Hats in the notation indicate that the recursion holds for operators in $D_i$,
\be
m_{\text{AdS}}(\I_n | \I_n) = \hat{m}_{\text{AdS}}(\I_n | \I_n)\,  \contact
\ee
stripped from the contact term.\footnote{This does not imply that the results can be upgraded from flat space, since individual $D_i \cdot D_j$'s do not commute in general.} Boundary condition is provided by $\hat{m}_{\text{AdS}}(\I_3 | \I_3) =1$, which is trivially obtained since $\mathcal{M}_{0,3}$ is a point.

The starting point is the correlator
\be\label{mAdS-n}
\hat{m}_{\text{AdS}}(\I_n | \I_n) = (-1)^{n-3}\oint_{\cap_{i=2}^{n-2} \Gamma_i} \frac{(z_{1,n-1}z_{n-1,n}z_{1n})^2}{(z_{12} z_{23} \cdots z_{n-1,n} z_{n1})^2} \bigwedge_{i=2}^{n-2} \frac{\d z_i}{2\pi i \, E_i}\, .
\ee
As before, in order to see which poles contribute, we substitute
\be
\hat{E}_i = E_i \prod_{\substack{k=1\\k\neq i}}^{n} z_{ik} = \sum_{\substack{j=1\\j\neq i}}^{n} D_{ij}^2 \prod_{\substack{k=1\\k\neq i,j}}^{n} z_{ik}\, ,
\ee
which yields
\be
\hat{m}_{\text{AdS}}(\I_n | \I_n) = \oint_{\cap_{i=2}^{n-2} \Gamma_i} \frac{z_{1,n-1}^2 \prod_{i=2}^{n-2} \prod_{j \neq i-1,i,i+1} z_{ij} }{ z_{12}\, z_{n-2,n-1}} \bigwedge_{i=2}^{n-2} \frac{\d z_i}{2\pi i \, \hat{E}_i}\, .
\ee
The contour of integration is defined with $\Gamma_i = \{|\hat{E}_i| = \varepsilon \}$. We will use the global residue theorem several times to massage the contour to a form that resembles the required recursion relation \eqref{recursion-relation}. As before, we will make use of
\be
\gamma_{21} = \{ |z_2 - z_1|  = \varepsilon \}, \qquad \gamma_{n-2,n-1} = \{ |z_{n-2} - z_{n-1}|  = \varepsilon \}\, .
\ee
We first write
\begin{align}\label{Gamma}
0 &= (\Gamma_2 + \gamma_{n-2,n-1}) \cap \Gamma_3 \cap \cdots \cap \Gamma_{n-3} \cap (\Gamma_{n-2} + \gamma_{21})\nn\\
&= \Gamma_2 \cap \Gamma_3 \cap \cdots \cap \Gamma_{n-3} \cap \Gamma_{n-2} \;+\; \Gamma_2 \cap \Gamma_3 \cap \cdots \cap \Gamma_{n-3} \cap  \gamma_{21}\nn\\
&\quad +   \gamma_{n-2,n-1} \cap \Gamma_3 \cap \cdots \cap \Gamma_{n-3} \cap \Gamma_{n-2} \;+\;  \underbrace{\gamma_{n-2,n-1} \cap \Gamma_3 \cap \cdots \cap \Gamma_{n-3} \cap \gamma_{21}}_{\tilde\Gamma}\, , 
\end{align}
where equality to zero should be understood as the appropriate homology statement. In the final equality the first term is original contour, while the second and third are the going to be responsible for the first two terms in the recursion \eqref{recursion-relation}, in analogy with what we have seen for $n=5$. The final term $\tilde\Gamma$ needs further manipulation. We use
\begin{align}
0 &= \gamma_{n-2,n-1} \cap (\Gamma_3 + \Gamma_{n-2}) \cap \Gamma_4 \cap \cdots \cap (\Gamma_{n-3} + \Gamma_2) \cap \gamma_{21}\nn\\
&= \tilde\Gamma  + \gamma_{21} \cap \gamma_{n-2,n-1} \cap \Gamma_4 \cap \cdots \cap \Gamma_{n-2} \;+\; \Gamma_2 \cap \Gamma_3 \cap \cdots \cap \Gamma_{n-4} \cap \gamma_{21} \cap \gamma_{n-2,n-1}\nn\\
&\quad- \Gamma_2 \cap \gamma_{n-2,n-1} \cap \Gamma_4 \cap \cdots \cap \Gamma_{n-4} \cap \gamma_{21} \cap \Gamma_{n-2} \, .
\end{align}
Recognizing that the last term looks structurally the same as $\tilde\Gamma$, we can repeat the same move until we arrive at
\be\label{Gamma-tilde}
\tilde{\Gamma} = - \sum_{k=2}^{n-3} \Gamma_2 \cap \Gamma_3 \cap \cdots \cap \Gamma_{k-1} \cap \underbrace{\gamma_{21} \cap \gamma_{n-2,n-1}}_{\text{replacing }\Gamma_{k}\cap \Gamma_{k+1}} \cap \Gamma_{k+2} \cap \cdots \cap \Gamma_{n-2}\, .
\ee
We will show that each term in this expansion leads exactly to the corresponding diagrams in the sum of the recursion \eqref{recursion-relation}.

Let us begin with the first contribution coming from the integral over $\Gamma_2 \cap \Gamma_3 \cap \cdots \cap \Gamma_{n-3} \cap  \gamma_{21}$. In parallel with the $n=5$ case, we can easily see that the contour $\gamma_{21}$ imposing $z_2 = z_1 + \delta x_2$ also implies that $z_{n-2}$ has to to approach $z_1$ at the same rate. Consider the following combination
\be
\sum_{i \in \{ 1,2,n-2\}} z_{i,n-2}\, E_i = z_{1,n-2} \frac{D_{12}^2}{\delta\, x_{12}} + z_{2,n-2} \frac{D_{12}^2}{\delta\, x_{21}} + \mathcal{O}(\delta^0) = D_{12}^2 + \mathcal{O}(\delta^0)\, ,
\ee
which thanks to the cancellation among the $\mathcal{O}(\delta^{-1})$ terms, stays finite. On the other hand, imposing $E_2=0$ from the $\Gamma_2$ part of the contour gives
\be
\sum_{i \in \{1,2,n-2\}} z_{i,n-2}\, E_i = z_{1,n-2} \left( \frac{D_{12}^2}{\delta\, x_{12}} + \frac{D_{12}^2}{z_{1,n-2}} + \ldots \right)\, ,
\ee
which implies that also $z_{n-2}$ needs to behave as $z_{n-2} = z_1 + \delta x_{n-2}$. As a matter of fact, repeating similar arguments one can show that
\be
z_i = z_1 + \delta x_i \qquad\text{for}\qquad i \in \{ 1, 2, n{-2}\} \cup S\, ,
\ee
with $x_1 = 0$ and a set of labels $S$ (which excludes $\{1,2,n{-}2\}$ and necessarily the gauge fixed punctures $n{-}1$ and $n$), is compatible with the above contour. However, only one of them will lead to a non-zero residue. To see this, let us count the degree of the pole in $\delta$ for all possible $S$. For this purpose we will consider the representation in \eqref{mAdS-n}. The volume form behaves as
\be
\bigwedge_{i=2}^{n-2} \d z_i = \mathcal{O}(\delta^{|S|+1})\, \d\delta
\ee
since exactly there are exactly $|S|$ factors of $\d x_{i\in S}$ and one of $\d x_{2}$ ($\d x_{n-2}$ is traded for $\d\delta$). The product of $E_i$ factors scales as
\be
\frac{1}{\prod_{i=2}^{n-2} E_i} = \mathcal{O}(\delta^{|S|+2})\, ,
\ee
because $E_i = \mathcal{O}(\delta^{-1})$ if $i \in \{2,n{-}2\} \cup S$ and is finite otherwise. The Jacobian factor only contains constants and hence does not scale. Finally, we are left with the factor of $(z_{12} z_{23} \cdots z_{n1})^{-2}$, which needs to scale at least as $\mathcal{O}(\delta^{-2|S|-4})$ for the integrand to have a pole. The only way this can be achieved is if all labels between $\{1,2\}$ and $\{n{-}2\}$ are in the set $S$, i.e.,
\be
S = \{ 3,4,\ldots,n{-}3\}\, .
\ee
All other degenerations do not have support on the residue around $\delta=0$.

The integral we would like to evaluate is given by
\be\label{first-contribution}
(-1)^{n-4}\oint_{\Gamma_2 \cap \Gamma_3 \cap \cdots \cap \Gamma_{n-3} \cap  \gamma_{21}} \frac{(z_{1,n-1}z_{n-1,n}z_{1n})^2}{(z_{12} z_{23} \cdots z_{n-1,n} z_{n1})^2} \bigwedge_{i=2}^{n-2} \frac{\d z_i}{2\pi i \, E_i}\, .
\ee
Making a change of variables to $(x_2, x_3, \ldots, x_{n-3},\delta)$ gives
\be
\bigwedge_{i=2}^{n-2} \d z_i = \delta^{n-4} x_{n-2,1}\, \d x_2 \wedge \d x_3 \wedge \cdots \wedge \d x_{n-3}\wedge \d\delta\, .
\ee
For $i=2,3,\ldots, n{-}2$ we find that the $E_i$ factors reduce to
\be
E_i = \frac{F_i}{\delta} + \mathcal{O}(\delta^0)\qquad \text{with}\qquad F_i =  \sum_{\substack{j=1\\ j \neq i}}^{n-2} \frac{D_{ij}^2}{x_i - x_j}\, .
\ee
In these variables the contour in \eqref{first-contribution} becomes $
\cap_{i=2}^{n-3} \{ | F_i| = \varepsilon \} \cap \{|\delta| = \varepsilon \}$.
Putting everything together, the integral in \eqref{first-contribution} reads
\be\label{first-contribution-2}
(-1)^{n-4}\oint_{\cap_{i=2}^{n-3} \{ | F_i| = \varepsilon \} \cap \{|\delta| = \varepsilon\} }  \frac{x_{n-2,1}}{(x_{12} x_{23} \cdots x_{n-3,n-2})^2} \bigwedge_{i=2}^{n-3} \frac{\d x_i}{2\pi i\, F_{i}} \wedge \left( \frac{1}{\delta} + \mathcal{O}(\delta^0)\right) \frac{ \d\delta}{2\pi i F_{n-2}}\, .
\ee
In order to simplify the $F_2$ term, let us evaluate the combination
\be\label{x-F-identity}
\sum_{i=1}^{n-2} x_{i1} F_i = \sum_{i=1}^{n-2} \sum_{\substack{j=1\\ j\neq i}}^{n-2} \frac{x_{ij}+x_{j1}}{x_{ij}} D_{ij}^2 = 2D_{12\ldots, n-2}^2 - \sum_{j=1}^{n-2} x_{j1} F_j\, ,
\ee
so the term on the left-hand side is equal to $D_{12\ldots, n-2}^2 = D_{n-1,n}^2$. However, on the support of the constraints $F_i = 0 $ for $i=2,3,\ldots,n{-}3$ this term becomes
\be
D_{n-1,n}^2 = x_{n-2,1} F_{n-2}\, .
\ee
Plugging this back into \eqref{first-contribution-2} and performing the trivial $\delta$ residue yields
\be\label{first-contribution-3}
\frac{(-1)^{n-4}}{D_{n-1,n}^2}\oint_{\cap_{i=2}^{n-3} \{ | F_i| = \varepsilon \} }  \frac{x_{n-2,1}^2}{(x_{12} x_{23} \cdots x_{n-3,n-2})^2} \bigwedge_{i=2}^{n-3} \frac{\d x_i}{2\pi i\, F_{i}}\, .
\ee
(The factor of $1/D_{n-1,n}^2$ can be commuted to the left since the factor $F_{n-2}$ commuted with all the remaining $F_i$'s.) We notice that this a gauge-fixed version of
\be
\frac{1}{D_{n-1,n}^2} \left( (-1)^{n-4} \oint_{\cap_{i=2}^{n-3} \{ | F_i| = \varepsilon \} }  \frac{(x_{1,n-2} x_{n-2,\ast} x_{\ast 1})^2}{(x_{12} x_{23} \cdots x_{n-3,n-2}\, x_{n-2,\ast}\, x_{1\ast})^2} \bigwedge_{i=2}^{n-3} \frac{\d x_i}{2\pi i\, F_{i}} \right)\, ,
\ee
with the fully $\SL(2,\C)$-covariant version of $F_i$, 
\be
F_i = \sum_{\substack{j=1\\ j\neq i}}^{n-2} \frac{D_{ij}^2}{x_i - x_j} + \frac{D_{i\ast}^2}{x_i - x_\ast}\, ,
\ee
where \eqref{first-contribution-3} can be obtained by fixing $x_\ast \to \infty$. By checking $\SL(2,\C)$-invariance of the resulting formula one finds that the conformal generator associated to the emergent puncture $x_\ast$ is
\be
D_\ast = D_{n-1} + D_n\, .
\ee
We have therefore confirmed that the this contribution gives
\be
\frac{1}{D_{n-1,n}^2} \hat{m}_{\text{AdS}}(12\cdots n{-}2,(n{-1},n) \,|\, 12\cdots n{-}2,(n{-1},n) )\, ,
\ee
which is indeed the first contribution to the recursion \eqref{recursion-relation}. By symmetry under the exchange of labels $(12\ldots n{-}2,n{-}1,n) \to (n{-}1,n{-}2,\ldots21n)$ one finds that the contour $\gamma_{n-2,n-1} \cap \Gamma_3 \cap \cdots \cap \Gamma_{n-3} \cap \Gamma_{n-2}$ from \eqref{Gamma} gives
\be
\frac{1}{D_{n1}^2} \hat{m}_{\text{AdS}}(2\cdots n{-}1,(n1) \,|\, 2\cdots n{-}1,(n1) )\, ,
\ee
which is the second term in the recursion relation.

It remains to study the contributions from contours $\Gamma_2 \cap \Gamma_3 \cap \cdots \cap \Gamma_{k-1} \cap \gamma_{21} \cap \gamma_{n-2,n-1} \cap \Gamma_{k+2} \cap \cdots \cap \Gamma_{n-2}$ for $k=2,3,\ldots,n{-}3$ coming from \eqref{Gamma-tilde}. By essentially identical arguments to those employed before one sees that on the support of $\gamma_{21}$ multiple punctures can collide with $z_1$ at some rate $\delta$, and likewise $\gamma_{n-2,n-1}$ allows another set to coalesce into $z_{n-1}$ at a rate $\epsilon$. Precisely one configuration has support on the corresponding residues around $\delta=\epsilon=0$, which can be parametrized by:
\be
z_i = \begin{dcases}z_1 + \delta x_i \qquad &\text{for} \qquad i=1,2,\ldots,k\, ,\\
	z_{n-1} + \epsilon y_i\qquad &\text{for} \qquad i=k{+}1,k{+}2,\ldots,n{-}1\, ,
\end{dcases}
\ee
with $x_1,y_{n-1} = 0$. The volume form becomes simply
\be
\bigwedge_{i=2}^{n-2} \d z_i = \delta^{k-2}\epsilon^{n-k-3}\, x_{k1}\, y_{k+1,n-1} \bigwedge_{i=2}^{k-1} \d x_i \wedge \d\delta \wedge \d\epsilon \wedge\!\!\! \bigwedge_{j=k+2}^{n-2} \d y_{j}
\ee
and once again we extract the leading behavior of the $E_i$ factors:
\be
E_i = \begin{dcases}\frac{F_i}{\delta} + \mathcal{O}(\delta^0)\quad  \text{with}\quad  F_i = \sum_{\substack{j=1\\ j \neq i}}^{k} \frac{D_{ij}^2}{x_i - x_j} \quad &\text{for} \quad i=1,2,\ldots,k\, ,\\
\frac{G_i}{\epsilon} + \mathcal{O}(\epsilon^0)\quad  \text{with}\quad  G_i = \sum_{j=k+1, j \neq i}^{n-1} \frac{D_{ij}^2}{y_i - y_j} \quad &\text{for} \quad i=k{+}1,\ldots,n{-}1\, .
\end{dcases}
\ee
With these assignments the integral over the aforementioned contour becomes
\begin{multline}
(-1)^{n-3}\oint_{\tilde{\Gamma}_{k,k+1}}  \frac{x_{k1}}{(x_{12}\, x_{23} \cdots x_{k-1,k})^2} \bigwedge_{i=2}^{k-1} \frac{\d x_i}{2\pi i\, F_{i}} \wedge \left( \frac{1}{\delta \epsilon} + \mathcal{O}(\delta^0,\epsilon^0)\right) \frac{ \d\delta \wedge \d\epsilon}{(2\pi i)^2 F_{k} G_{k+1}}\\
 \wedge \frac{y_{k+1,n-1}}{(y_{k+1,k+2}\, y_{k+2,k+3} \cdots y_{n-2,n-1})^2} \bigwedge_{j=k+1}^{n-2} \frac{\d y_j}{2\pi i\, G_{j}}\,,\label{third-contribution}
\end{multline}
where the contour of integration is given by
\be
\tilde{\Gamma}_{k,k+1} = \bigcap_{i=2}^{k-1} \{ |F_i| = \varepsilon \}\cap \{ |\delta| = \varepsilon\} \cap \{ |\epsilon| = \varepsilon\} \cap\!\!\! \bigcap_{j=k+2}^{n-2} \{ |G_j| = \varepsilon \}\, .
\ee
Repeating the derivation used in \eqref{x-F-identity} we find on the support of the above contour
\be
D_{12\ldots k}^2 = \sum_{i=1}^{k} x_{i1} F_a = x_{k1} F_k
\ee
and
\be
D_{k+1, k+2, \ldots, n-1}^2 = \sum_{j=k+1}^{n-1} y_{i,n-1} G_j = y_{k+1,n-1}G_{k+1}\, .
\ee
With these substitutions into the integral \eqref{third-contribution}, performing the trivial residues around $\delta=\epsilon=0$ and restoring $\SL(2,\C)$ invariance one finds that the answer factors into two integrals
\begin{multline}
\frac{1}{D_{12\ldots k}^2\, D_{k+1, k+2, \ldots, n-1}^2} (-1)^{k-2} \oint_{\cap_{i=2}^{k-1} \{ | F_i| = \varepsilon \} }  \frac{(x_{1k} x_{k\ast} x_{1\ast})^2}{(x_{12} x_{23} \cdots x_{k-1,k}\, x_{k\ast}\, x_{\ast 1})^2} \bigwedge_{i=2}^{k-1} \frac{\d x_i}{2\pi i\, F_{i}}\\
\times(-1)^{n-k-3} \oint_{\cap_{j=k+2}^{n-2} \{ | G_j| = \varepsilon \} }  \frac{(y_{k+1,n-1} y_{k+1\star} y_{n-1\star})^2}{(y_{k+1,k+2} y_{k+2,k+3} \cdots y_{n-2,n-1}\, y_{n-1,\star}\, y_{\star,k+1})^2} \bigwedge_{j=k+1}^{n-2} \frac{\d y_j}{2\pi i\, G_{j}} \, .
\end{multline}
Here once again the factors of $1/D_e^2$ can be placed up front since the positions of scattering equations were immaterial to begin with. By studying transformation properties of $F_i$ and $G_j$'s one finds that conformal generators associated to the punctures $x_\ast$ and $y_\star$ are respectively
\be
D_\ast = D_{12\ldots k}, \qquad D_\star = D_{k+1, k+2, \ldots, n-1}\, .
\ee
We have thus found that the above product of integrals equals
\begin{multline}
\frac{1}{D_{12\dots k}^2\, D_{k+1, k+2, \dots, n-1}^2} \hat{m}_{\text{AdS}}( 12\cdots k (k{+}1,\dots,n)\,|\, 12\dots k (k{+}1,\dots,n) )\\
\times \hat{m}_{\text{AdS}}(k{+}1,k{+}2,\dots,n{-}1(n1\dots k)\,|\, k{+}1,k{+}2,\dots,n{-}1(n1\dots k))
\end{multline}
for any $k=2,3,\ldots,n{-}3$, which are indeed the remaining terms in the recursion relation \eqref{recursion-relation}. This concludes the proof.

\subsubsection{General permutations}

The proof in the general case, $\sigma \neq \tau$, proceeds by repeating the same steps. Setting $\sigma = \I_n$ without loss of generality, the only difference is that the Parke-Taylor factor corresponding to $\tau$ induces only a \emph{subset} of poles of those present in the $\tau = \I_n$ case. Hence only a subset of trivalent degenerations will contribute non-zero residues. It is a matter of bookkeeping to describe which graphs do contribute. Since this is no different to the flat-space case, and we already proved that all planar degenerations are obtained correctly for operator-valued scattering equations, it necessarily means that $\hat{m}_{\text{AdS}}(\sigma | \tau)$ must obey the same recursion relations as its flat-space counterpart \cite{Mafra:2016ltu}. To state them, let us fix $n$ to the final slot of both $\sigma = (\hat\sigma, n)$ and $\tau = (\hat{\tau},n)$, where $\hat\sigma$ and $\hat{\tau}$ are permutations of the remaining labels $\{1,2,\ldots,n{-}1 \}$. Then we have
\begin{multline}\label{recursion-relation-off-diagonal}
\hat{m}_{\text{AdS}}(\hat{\sigma},n | \hat{\tau}, n) = \sum_{\hat{\sigma} = \hat{\sigma}_1 \hat{\sigma}_2} \sum_{\hat{\tau} = \hat{\tau}_1 \hat{\tau}_2} \frac{1}{D_{\hat\sigma_1}^2 D_{\hat\sigma_2}^2} \bigg( \eta_{\hat\sigma_1,\hat{\tau}_1}\, \hat{m}_{\text{AdS}}(\hat{\sigma}_1,(\hat{\sigma}_2,n) \,|\, \hat{\tau}_1, (\hat{\tau}_2,n)) \\
\times \hat{m}_{\text{AdS}}(\hat{\sigma}_2,(n,\hat{\sigma}_1) \,|\, \hat{\tau}_2, (n,\hat{\tau}_1)) - (\hat\sigma_1 \leftrightarrow \hat{\sigma}_2) \bigg)\, .
\end{multline}
The two sums go over all ways of deconcatenating $\sigma$ and $\tau$ into smaller words, with each $\hat{\sigma}_i$ and $\hat{\tau}_i$ having at least one element. One needs to supply a boundary condition for when $\hat{\sigma}$ and $\hat{\tau}$ have a single element each, say $\hat{\sigma}=(a)$ and $\hat{\tau}=(b)$, given by
\be\label{boundary-condition}
\hat{m}_{\text{AdS}}(an | bn) = \delta_{ab}\, D_{a}^2\,,
\ee
where for the purposes of the recursion we treat $D_i^2$ as a non-zero formal variable, which always cancels out from the final expression.
Inside the sum $\eta_{\hat\sigma_1,\hat{\tau}_1}$ equals to $1$ when the two words consist of the same sets of labels (which implies an analogous statement for $\hat{\sigma}_2$ and $\hat\tau_2$'s) and $0$ otherwise. In particular, it means that $D_{\hat{\sigma}_i}^2 = D_{\hat{\tau}_i}^2$.

It is straightforward to see that when $\sigma = \tau$, the above recursion boils down to \eqref{recursion-relation}. The condition $\eta_{\hat\sigma_1,\hat{\tau}_1}$ means that all $\hat\sigma_i = \hat\tau_i$. Moreover, the second term in the brackets corresponding to antisymmetrization $(\hat\sigma_1 \leftrightarrow \hat{\sigma}_2)$ never contributes, since $\hat{\sigma}_{1/2}$ can never have the same labels as $\hat{\tau}_{2/1}$. The sum has exactly $n{-}2$ terms corresponding to partitioning $\hat{\sigma} = \hat{\tau}$ into two subwords with at least one element each. Noting the boundary condition \eqref{boundary-condition}, the edge cases when either $\hat{\sigma}$ or $\hat{\tau}$ have a single element are the first two terms of \eqref{recursion-relation}. The remaining $n{-}4$ terms are precisely those from the final sum in \eqref{recursion-relation}.

As an example of the evaluation of \eqref{recursion-relation-off-diagonal} with $\sigma \neq \tau$, let us consider $\sigma = (12345)$ and $\tau = (21435)$. There is only one way of deconcatenating $\hat{\sigma} = (1234)$ and $\hat{\tau} = (2143)$ into words that contain the same subsets of labels, which is given by
\be
\hat{\sigma}_1 = (12),\qquad \hat{\sigma}_2 = (34), \qquad \hat{\tau}_1 = (21), \qquad \hat{\tau}_2 = (43)\,.
\ee
With only the first term in the sum having support we find
\be\label{mAdS-12345}
\hat{m}_{\text{AdS}}(12345|21435) = \frac{1}{D_{12}^2 D_{34}^2} \hat{m}_{\text{AdS}}(12(345)|21(345))\, \hat{m}_{\text{AdS}}(34(512)|43(512))\,.
\ee
Let us proceed with the first factor on the right-hand side, which has $\hat{\sigma} = (12)$ and $\hat{\tau} = (21)$, and therefore a single compatible deconcatenation:
\be
\hat{\sigma}_1 = (1),\qquad \hat{\sigma}_2 = (2), \qquad \hat{\tau}_1 = (2), \qquad \hat{\tau}_2 = (1)\,.
\ee
Only the second term in the sum in \eqref{recursion-relation-off-diagonal} contributes, giving
\be
\hat{m}_{\text{AdS}}(12(345)|21(345)) = - \frac{1}{D_{1}^2 D_{2}^2} \hat{m}_{\text{AdS}}(1(2345)|1(3452))\, \hat{m}_{\text{AdS}}(2(3451)|2(1345)) = -1\,,
\ee
where we used the boundary condition \eqref{boundary-condition}. The second contribution in \eqref{mAdS-12345} also evaluates to $-1$ by symmetry. We therefore have
\be
\hat{m}_{\text{AdS}}(12345|21435) = \frac{1}{D_{12}^2 D_{34}^2}\,,
\ee
in agreement with \eqref{m-AdS-result} since $w(12345|21435)=3$.

\subsection{Comparison with Witten diagrams}

We now show that the formula \eqref{m-AdS-result} reproduces the results of perturbation theory in AdS. This follows rather straightforwardly from the following ``intertwining'' property of the bulk-to-boundary propagator.

\paragraph{Bulk-to-boundary propagator as intertwiner.} Let us  first consider a single bulk-to-boundary propagator $1/(-2X\cdot P)^{\Delta}$. Since the expression is manifestly invariant under the SO$(2,d)$ transformations, it satisfies
\be
D_P^{a}\frac{1}{(-2X\cdot P)^{\Delta}}=-D_X^{a}\frac{1}{(-2X\cdot P)^{\Delta}}\,,
\ee
where $D_P^{a}$ and $D_X^{a}$ are given by
\begin{align}
&D_P^{a}=D_P^{[AB]}=P^{A}\frac{\partial}{\partial P_{B}}-P^{B}\frac{\partial}{\partial P_{A}}\,,\qquad D_X^{a}=D_X^{[AB]}=X^{A}\frac{\partial}{\partial X_{B}}-X^{B}\frac{\partial}{\partial X_{A}}\,.
\end{align}
Using this, we can rewrite the action of the Casimir $D_P\cdot D_P$ into the action of the Laplacian in AdS, $\Box_{X}\equiv D_X\cdot D_X/2$:
\be
D_P\cdot D_P \left(\frac{1}{(-2X\cdot P)^{\Delta}}\right)=2\Box_{X}\left(\frac{1}{(-2X\cdot P)^{\Delta}}\right)\,.
\ee

A similar relation holds also for a product of the bulk-to-boundary propagators. Namely, using
\be
\left(D_X^{a}+\sum_{i=1}^{n}D_{P_i}^{a}\right)\left(\prod_{i=1}^{n}\frac{1}{(-2X\cdot P_i)^{\Delta_i}}\right)=0\,,
\ee
we can derive the relation
\be\label{eq:intertwining}
D_{12\ldots n}^2 \left(\prod_{i=1}^{n}\frac{1}{(-2X\cdot P_i)^{\Delta_i}}\right)=2\Box_{X}\left(\prod_{i=1}^{n}\frac{1}{(-2X\cdot P_i)^{\Delta_i}}\right)\,,
\ee
with $D^{2}_{12\ldots n}\equiv (D_{P_1}+\cdots +D_{P_n})^2$.

\begin{figure}
 	\centering
 	\includegraphics{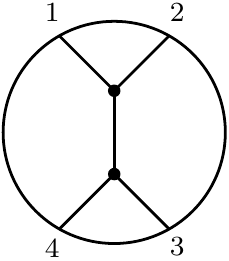}\hspace{5em}
 	\includegraphics{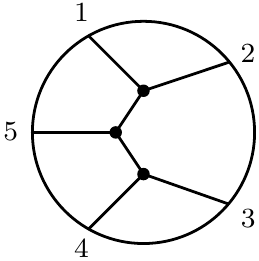}
 	\caption{\label{fig:diagrams}Two examples of Witten diagrams corresponding to the expressions \eqref{eq:simple4pt} and \eqref{eq:simple5pt} respectively.}
 \end{figure}

\paragraph{Proof of equivalence with Witten diagrams.} Using the intertwining relation \eqref{eq:intertwining} and the integral representation for the contact diagram
\be
\mathcal{C}=\int_{\text{AdS}_{d+1}} \!\!\! \d X\ \prod_{i=1}^{n}\frac{1}{(-2X\cdot P_{i})^{d}}\,,
\ee
 we can rewrite every factor $1/D_{e}^2$ appearing in the formula \eqref{m-AdS-result} into the insertion of $1/\Box_{X}$. For instance, the term $(1/D_{12}^2)\mathcal{C}$ in the four-point function can be expressed as
 \be\label{eq:simple4pt}
\frac{1}{D_{12}^2}\mathcal{C}=\frac{1}{2}\int _{\text{AdS}_{d+1}} \d X\  \frac{1}{(-2X\cdot P_3)^d(-2X\cdot P_4)^{d}}\frac{1}{\Box_{X}}\left[\frac{1}{(-2X\cdot P_1)^d(-2X\cdot P_2)^{d}}\right]\,. 
 \ee
 Since the inverse Laplacian $1/\Box_{X}$ is nothing but the propagator of a massless particle in AdS, the right hand side of \eqref{eq:simple4pt} coincides with the s-channel exchange Witten diagram shown in Figure~\ref{fig:diagrams} (left). Similarly, the term $(1/D_{12}^2D_{34}^2) \mathcal{C}$ in the five-point function can be expressed as
 \begin{multline}
 \frac{1}{D_{12}^2D_{34}^2}\mathcal{C}=\frac{1}{4} \int _{\text{AdS}_{d+1}}\!\!\!\!\!\d X\  \frac{1}{(-2X\cdot P_5)^d}\frac{1}{\Box_{X}}\left[\frac{1}{(-2X\cdot P_3)^d(-2X\cdot P_4)^{d}}\right]\\
 \times\frac{1}{\Box_{X}}\left[\frac{1}{(-2X\cdot P_1)^d(-2X\cdot P_2)^{d}}\right]\,,\label{eq:simple5pt}
 \end{multline}
 and reproduces the Witten diagram given in Figure~\ref{fig:diagrams} (right).

Based on this argument, one can rewrite the right hand side of \eqref{m-AdS-result} as a sum over trivalent tree-level Witten diagrams that are planar with respect to the permutations $\sigma$ and $\tau$. As is well-known in flat space \cite{Cachazo:2013hca,Cachazo:2013iea}, this sum coincides precisely with the result of perturbation theory in the bi-adjoint scalar theory. This gives a formal proof of equivalence between our formalism and the massless bi-adjoint scalar theory in AdS.

 \paragraph{Spectral representation of correlation functions.} The expressions obtained by the argument above, \eqref{eq:simple4pt} and \eqref{eq:simple5pt}, are not necessarily useful for practical purposes since they involve the inverse of the AdS Laplacian $1/\Box_{X}$. Below we explain how to rewrite them into a form more suited for extracting the conformal data.
 
 The first step of the rewriting is to insert as many AdS delta functions as the number of $1/\Box$'s. For instance, \eqref{eq:simple5pt} can be rewritten as
 \begin{multline}
 \frac{4}{D_{12}^2D_{34}^2}\mathcal{C}=\int _{\text{AdS}_{d+1}} \frac{\d X_1 \d X_2 \d X_3}{(-2X_1\cdot P_1)^d(-2X_1\cdot P_2)^{d}(-2X_2\cdot P_3)^d(-2X_2\cdot P_4)^{d}(-2X_3\cdot P_5)^d}\\
 \times \frac{1}{\Box_{X_1}\Box_{X_2}}\left[\delta^{d+1}(X_1-X_3)\delta^{d+1}(X_2-X_3)\right]\,.
 \end{multline}
 Here, we also integrated twice by parts in order to move the $1/\Box$ on the delta functions.
In the next step, we use the spectral representation of the delta function \cite{Penedones:2007ns,Penedones:2010ue}
 \be
 \delta^{d+1}(X-Y)=\int_{-\infty}^{\infty} \d\nu\,\, \Omega_{\nu} (X,Y)\,,
 \ee
 where $\Omega_{\nu} (X,Y)$ is a harmonic function in AdS, which satisfies\footnote{See \cite{Cornalba:2007fs} for its explicit expression.}
 \begin{align}\label{eq:laplaceomega}
    \Box_{X} \Omega_{\nu}(X,Y)=-\left(\nu^2 +\frac{d^2}{4}\right)\Omega_{\nu} (X,Y)\,.
\end{align}
We then get
  \begin{multline}
 \frac{4}{D_{12}^2D_{34}^2}\mathcal{C}=\int _{\text{AdS}_{d+1}} \frac{\d X_1 \d X_2 \d X_3}{(-2X_1\cdot P_1)^d(-2X_1\cdot P_2)^{d}(-2X_2\cdot P_3)^d(-2X_2\cdot P_4)^{d}(-2X_3\cdot P_5)^d}\\
 \times \frac{1}{\Box_{X_1}\Box_{X_2}}\int \d\nu_1 \d\nu_2 \,\Omega_{\nu_1}(X_1,X_3)\Omega_{\nu_2}(X_2,X_3)\,,
 \end{multline}
 which can be further rewritten using \eqref{eq:laplaceomega} as
\begin{multline}
 \frac{4}{D_{12}^2D_{34}^2}\mathcal{C}=\int _{\text{AdS}_{d+1}} \frac{\d X_1 \d X_2 \d X_3}{(-2X_1\cdot P_1)^d(-2X_1\cdot P_2)^{d}(-2X_2\cdot P_3)^d(-2X_2\cdot P_4)^{d}(-2X_3\cdot P_5)^d}\\
 \times\int \d\nu_1 \d\nu_2 \,\frac{\Omega_{\nu_1}(X_1,X_3)}{\nu_1^2+\frac{d^2}{4}}\frac{\Omega_{\nu_2}(X_2,X_3)}{\nu_2^2+\frac{d^2}{4}}\,.\label{eq:bulktobulkspectral}
 \end{multline}
Note that the expression that appear on the second line,
\be
G(X,Y)\equiv \int \d\nu \frac{\Omega_{\nu}(X,Y)}{\nu^2 +\frac{d^2}{4}}\,,
\ee
coincides with the spectral representation of the bulk-to-bulk propagator of massless particle known in the literature \cite{Penedones:2010ue}. This also confirms the previous assertion that $(1/D_{12}^2D_{34}^2)\mathcal{C}$ reproduces the corresponding Witten diagram.

 The last step of the rewriting is to use the {\it split representation} for the bulk-to-bulk propagator of massless particle $G(X,Y)$ \cite{Penedones:2010ue}
 \be
 G (X,Y)=\int\frac{\d\nu}{\pi}\frac{\nu^2}{\nu^2+\frac{d^2}{4}}\int \frac{\d\tilde{P}\, \mathcal{N}_{\frac{d}{2}+i\nu}\mathcal{N}_{\frac{d}{2}-i\nu}}{(-2X\cdot \tilde{P})^{\frac{d}{2}+i\nu}(-2Y\cdot \tilde{P})^{\frac{d}{2}-i\nu}} \,,
 \ee
where $P$ is a boundary point and the normalization constant $\mathcal{N}_{\Delta}$ is given by
\begin{align}
    \mathcal{N}_{\Delta}=\frac{\Gamma (\Delta)}{2\pi^{d/2}\Gamma (\Delta -\frac{d}{2}+1)}\,.
\end{align}
Applying this to \eqref{eq:bulktobulkspectral}, we obtain the following expression
\begin{align}
 \frac{4}{D_{12}^2D_{34}^2}\mathcal{C}=&\int \frac{\d\nu_1}{\pi}\frac{\d\nu_2}{\pi} \d\tilde{P}_1 \d\tilde{P}_2(\cdots)\\
 &\times \int _{\text{AdS}_{d+1}} \frac{\d X_1 \d X_2 \d X_3}{(-2X_1\cdot P_1)^d(-2X_1\cdot P_2)^{d}(-2X_2\cdot P_3)^d(-2X_2\cdot P_4)^{d}(-2X_3\cdot P_5)^d}\nonumber\\
 &\times \frac{1}{(-2X_1\cdot \tilde{P}_1)^{\frac{d}{2}+i\nu_1}(-2X_3\cdot \tilde{P}_1)^{\frac{d}{2}-i\nu_1}(-2X_2\cdot \tilde{P}_2)^{\frac{d}{2}+i\nu_2}(-2X_3\cdot \tilde{P}_2)^{\frac{d}{2}-i\nu_2}}\,.\nonumber
\end{align}
It is then straightforward to perform the integrals of bulk points $X_j$ using the formula \cite{Muck:1998rr}
\be
\int_{\text{AdS}_{d+1}}  \frac{\d X}{(-2X\cdot P_1)^{\Delta_1}(-2X\cdot P_2)^{\Delta_2}(-2X\cdot P_3)^{\Delta_3}}=\frac{B_{\Delta_1,\Delta_2,\Delta_3}}{(P_{12})^{\Delta_{12|3}}(P_{23})^{\Delta_{23|1}}(P_{31})^{\Delta_{31|2}}}\,,
\ee
with $\Delta_{ij|k}\equiv (\Delta_i+\Delta_j-\Delta_k)/2$, $P_{ij}\equiv-2P_i \cdot P_j$ and
\be
B_{\Delta_1,\Delta_2,\Delta_3}\equiv \frac{\pi^{\frac{d}{2}}}{2}\Gamma\left(\tfrac{\Delta_1+\Delta_2+\Delta_3-d}{2}\right)\frac{\Gamma\left(\frac{\Delta_1+\Delta_2-\Delta_3}{2}\right)\Gamma\left(\frac{\Delta_1-\Delta_2+\Delta_3}{2}\right)\Gamma\left(\frac{-\Delta_1+\Delta_2+\Delta_3}{2}\right)}{\Gamma\left(\Delta_1\right)\Gamma\left(\Delta_2\right)\Gamma\left(\Delta_3\right)}\,.
\ee
We are then left with the conformal integrals of the boundary points $\tilde{P}_i$ and the integrals of the spectral parameters $\nu_j$. From such expressions, one can extract the conformal data (such as the conformal dimensions and the structure constants) using various techniques developed in the literature, see for instance \cite{Liu:2018jhs,Karateev:2018oml}. It would be interesting to work them out explicitly for lower-point examples, but we leave it for a future investigation. 
\section{\label{sec:eigenfunctions}Eigenfunctions of the AdS scattering equations}
\subsection{Scattering equation as interpolating Casimir}
Let us recall the integral for the color-ordered four-point function
\be\label{eq:recap}
m_{\rm AdS}(\mathbb{I}_4 |\mathbb{I}_4)=\oint_{\Gamma_2} \frac{z_{13}^2z_{24}}{z_{12}z_{23}}\frac{dz_2}{2\pi i \hat{E}_2}\mathcal{C}\,,
\ee
with
\be\label{eq:4ptscatteringcas}
\hat{E}_2 =z_{23}z_{24}D_{12}^2+z_{21}z_{24}D^{2}_{23}+z_{21}z_{23}D^{2}_{24}\,.
\ee
In Section \ref{sec:correlators}, we evaluated this integral by the residues at the poles $z_2=z_{1}$ and $z_2=z_3$. A possible alternative would be to evaluate the integral directly at $\hat{E}_2=0$. For this purpose, one needs to decompose the integrand into eigenfunctions of $\hat{E}_2$ and replace $\hat{E}_2$ with its eigenvalues $e_2 (\nu)$, 
\be\label{eq:eigenequation}
\hat{E}_2 \varphi_{\nu} =-z_{13}z_{24}e_2 (\nu) \varphi_{\nu}\,.
\ee
Here $\varphi_{\nu}$ and $e_2 (\nu)$ are an eigenfunction and an eigenvalue, and $\nu$ is a quantum number which distinguishes different eigenfunctions. We also introduced a prefactor $-z_{13}z_{24}$ on the right hand side for later convenience.
After this replacement, \eqref{eq:recap} reduces to an integral which only involves $c$-numbers, and we can try to evaluate it using the standard residue theorem. Of course, we need to work out various details---such as a complete basis of eigenfunctions and a decomposition of the contact diagram into eigenfunctions---in order to substantiate this idea. In what follows, we take an initial step toward such a direction: We analyze the eigenvalue equation \eqref{eq:eigenequation} for the four-point functions in AdS$_{d+1}$, and rewrite it into the $BC_2$ Inozemtsev model \cite{inozemtsev1989lax,takemura2003inozemtsev}.

 Before analyzing \eqref{eq:eigenequation}, let us emphasize an important property of the scattering equation $\hat{E}$ that it interpolates conformal Casimirs in different channels. This is evident in the expression \eqref{eq:4ptscatteringcas}: At $z_2=z_{1}$, $\hat{E}_2$ becomes proportional to the conformal Casimir in the $s$-channel $D_{12}^2$, while it is proportional to the Casimirs in the $t$- and $u$-channels at $z_2=z_3$ and $z_2=z_4$. For other values of $z_2$, it gives a differential operator which interpolates between the Casimirs in those three channels. For this reason, we call the solutions to \eqref{eq:eigenequation} generalized conformal partial waves. This interpolation property is true also for higher-point functions, as can be verified straightforwardly from the definition of the scattering equation \eqref{eq:scattering-equations}. 
\subsection{Eigenfunctions and relation to Inozemtsev model} We now analyze the eigenvalue equation \eqref{eq:eigenequation} in AdS$_{d+1}$. Below we allow the four operators to have arbitrary conformal dimensions $\Delta_j$ $(j=1,\ldots, 4)$. This is a slight generalization of the setup in the main text, in which all the operators had dimension $\Delta_j=d$.

Let us first recall the derivation of standard conformal Casimir equations. The first step is to factor out simple kinematical dependences from the eigenfunction so that the rest depends only on the conformal cross ratios $x$ and $\bar{x}$,
\be
\frac{P_{12}P_{34}}{P_{13}P_{24}}=x\bar{x}\,,\qquad \frac{P_{14}P_{23}}{P_{13}P_{24}}=(1-x)(1-\bar{x})\,,
\ee
with $P_{ij}\equiv -2 P_i\cdot P_j$.
 For the $s$-channel conformal Casimir equation, a convenient choice is  
\be\label{eq:schanneldecomp}
\text{$s$-channel:} \qquad \frac{1}{P_{12}^{(\Delta_1+\Delta_2)/2}P_{34}^{(\Delta_3+\Delta_4)/2}}\left(\frac{P_{24}}{P_{14}}\right)^{\delta_{12}}\left(\frac{P_{14}}{P_{13}}\right)^{\delta_{34}} g_s (x,\bar{x})\,,
\ee
with $\delta_{ij}\equiv (\Delta_i-\Delta_j)/2$, while analogues for the $t$- and the $u$-channels read
\begin{align}
\text{$t$-channel:} & \qquad  \frac{1}{P_{23}^{(\Delta_2+\Delta_3)/2}P_{14}^{(\Delta_1+\Delta_4)/2}}\left(\frac{P_{24}}{P_{34}}\right)^{\delta_{32}}\left(\frac{P_{34}}{P_{13}}\right)^{\delta_{14}} g_t (x,\bar{x})\,,\label{eq:tchanneldecomp}\\
\text{$u$-channel:}& \qquad  \frac{1}{P_{24}^{(\Delta_2+\Delta_4)/2}P_{13}^{(\Delta_1+\Delta_3)/2}}\left(\frac{P_{12}}{P_{14}}\right)^{\delta_{42}}\left(\frac{P_{14}}{P_{34}}\right)^{\delta_{31}} g_u (x,\bar{x})\,.\label{eq:uchanneldecomp}
\end{align}
These representations allow us to translate the actions of the conformal Casimirs into differential operators acting on $g_{s,t,u}(x,\bar{x})$. For instance, the $s$-channel Casimir $-D_{12}^2/4$ can be translated to the following differential operator acting on $g_{s}(x,\bar{x})$ \cite{Dolan:2003hv}:
\be\label{eq:schannelcasimirdif}
\mathcal{D}_s =\mathbb{D}_x +\mathbb{D}_{\bar{x}}+(d-2)\frac{x\bar{x}}{x-\bar{x}}\left((1-x)\partial_x-(1-\bar{x})\partial_{\bar{x}}\right)\,,
\ee
with
\be
\mathbb{D}_x\equiv x^2 (1-x)\partial_{x}^2-(\delta_{21}+\delta_{34}+1)x^2 \partial_{x}-\delta_{21}\delta_{34}x\,.
\ee
The expressions for the actions of $-D_{23}^2/4$ on $g_t$ and $-D_{24}^2/4$ on $g_u$ can be obtained by performing the following replacements of the indices and the cross ratios to \eqref{eq:schannelcasimirdif}:
\begin{align}
\text{$t$-channel:}&  \qquad 1\leftrightarrow 3\,, \quad x\to 1-x\,,\quad \bar{x}\to 1-\bar{x}\,,\\
\text{$u$-channel:}&  \qquad 1\leftrightarrow 4\,, \quad x\to -\frac{1}{x}\,,\quad \bar{x}\to -\frac{1}{\bar{x}}\,.
\end{align}
These differential equations are often called Casimir differential equations.

Alternatively, we can act the $t$- and $u$-channel Casimirs $-D_{23}^2/4$ and $-D_{24}^2/4$ on the $s$-channel expression \eqref{eq:schanneldecomp} and express them as differential operators acting on $g_s$. The actions can be read off straightforwardly from the Casimir differential equations for the $t$- and $u$-channels, using the fact that the ratios between the kinematical prefactors in \eqref{eq:schanneldecomp}-\eqref{eq:uchanneldecomp} are functions of the cross ratios. Using such expressions, we can rewrite the eigenvalue equation for the scattering equation \eqref{eq:eigenequation} into a differential equation. After fixing the worldsheet SL$(2,\mathbb{C})$ redundancy by setting
\be
z_1=0,\quad z_2=z,\quad z_3=1,\quad z_4=\infty\,,
\ee
the result reads
\be
\mathcal{D}_{\hat{E}_2} f_{\nu} (x,\bar{x})=\left(e_2 (\nu)+\delta e_2\right)f_{\nu} (x,\bar{x})\,,
\ee
where $f_{\nu}(x,\bar{x})$ and $\delta e_2$ are given by
\begin{align}
&\varphi_{\nu}=\frac{1}{P_{12}^{(\Delta_1+\Delta_2)/2}P_{34}^{(\Delta_3+\Delta_4)/2}}\left(\frac{P_{24}}{P_{14}}\right)^{\delta_{12}}\left(\frac{P_{14}}{P_{13}}\right)^{\delta_{34}} (x\bar{x})^{\frac{\Delta_1+\Delta_2}{2}} f_{\nu}(x,\bar{x})\,,\\
&\delta e_2 =\frac{2(2z-1)}{3}\sum_{j=1}^{4}\Delta_j (\Delta_j-d)\,.
\end{align}
The differential operator $\mathcal{D}_{\hat{E}_2}$ is defined by
\be\label{eq:differentialoperator}
\begin{aligned}
\mathcal{D}_{\hat{E}_2}= & -(y_x)^2\left[\partial_{x}^2+\left(\frac{2 {\sf a}}{x-\bar{x}}+{\sf b}(x)\right)\partial_x\right] -(y_{\bar{x}})^2\left[\partial_{\bar{x}}^2+\left(\frac{2 {\sf a}}{\bar{x}-x}+{\sf b}(\bar{x})\right)\partial_{\bar{x}}\right]\\
&-2\Delta_2 (\Delta_1+\Delta_2+\Delta_3-\Delta_4)(x+\bar{x})+{\sf d}
\end{aligned}
\ee
with
\begin{align}
&y_t\equiv  \sqrt{t (t-1)(t-z)}\,,\qquad {\sf a}\equiv \frac{d-2}{2}\,,\qquad {\sf b}(x)\equiv \frac{\eta_1+\frac{1}{2}}{x}+\frac{\eta_2+\frac{1}{2}}{x-1}+\frac{\eta_3+\frac{1}{2}}{x-z}  \,,\nonumber\\
 &{\sf d}\equiv 2 \left[(\eta_1+\eta_2)^2 e_3+(\eta_2+\eta_3)^2 e_1+(\eta_3+\eta_1)^2 e_2\right]-2(d-2)\sum_{j=1}^{3} e_j \eta_j\,,\\
 &\eta_1\equiv \frac{1+\Delta_1+\Delta_2-\Delta_3-\Delta_4}{2}\,,\quad \eta_2\equiv \frac{1-\Delta_1+\Delta_2+\Delta_3-\Delta_4}{2}\,,\quad \eta_3\equiv \frac{1-2d+\sum_{j=1}^{4}\Delta_j}{2}\,,\nonumber\\
 &e_1\equiv -\frac{1+z}{3}\,,\qquad e_2\equiv \frac{2-z}{3}\,,\qquad e_3\equiv \frac{2z-1}{3}\,, \qquad (e_1+e_2+e_3=0)\,.\nonumber
\end{align}

Now, the crucial observation is that the differential operator \eqref{eq:differentialoperator} takes exactly the same form as the Hamiltonian $\hat{H}$ given in section 4 of \cite{takemura2003inozemtsev} if we shift of the coordinates $x\to x+\frac{1+z}{3}$ and $\bar{x}\to \bar{x}+\frac{1+z}{3}$. In that paper, $\hat{H}$ was obtained by performing a similarity transformation to the Hamiltonian of the $BC_2$ Inozemtsev model \cite{,takemura2003inozemtsev}, which is an elliptic deformation of the $BC_2$ Calogero-Sutherland model. Undoing the similarity transformation amounts to a further redefinition of the eigenfunction $f_{\nu}(x,\bar{x})$
\be
f_{\nu}(x,\bar{x})=\Phi (x,\bar{x})^{-1}\, \psi_{\nu}\,,
\ee
with
\be
\Phi (x,\bar{x})\equiv (x-\bar{x})^{{\sf a}}\prod_{j=1}^{3}(x\bar{x})^{\eta_1/2}\left((x-1)(\bar{x}-1)\right)^{\eta_2/2}\left((x-z)(\bar{x}-z)\right)^{\eta_3/2}\,.
\ee
After this redefinition, the eigenvalue equation becomes the Schr{\"o}dinger equation for the $BC_2$ Inozemtsev model,
\be
H\,\psi_{\nu}(u_1,u_2)=\left(e_2 (\nu)+\delta e_2\right)\psi_{\nu}(u_1,u_2)\,,
\ee
with
\be\label{eq:hamiltonianino}
\begin{aligned}
H \equiv &-\partial_{u_1}^2-\partial_{u_2}^2 +2{\sf a}({\sf a}-1)\left[ \wp (u_1-u_2)+\wp (u_1+u_2)\right]\\
&+\sum_{j=1}^{3}\eta_j(\eta_j-1)\left[\wp (u_1+\omega_j)+\wp (u_2+\omega_j)\right]\,,
\end{aligned}
\ee
Here $\wp (u)$'s are the Weierstrass $\wp$-functions associated with the elliptic curve\footnote{Thus, the two invariants ${\sf g}_2$ and ${\sf g}_3$ take the form
\be
{\sf g}_2 = \frac{4}{3}(1-z+z^2)\,,\qquad {\sf g}_3=\frac{4}{27}\left(2-3z-3z^2+2z^3\right)\,. 
\ee}
\be\label{eq:deftorus}
y^2 =x (x-1)(x-z)\,,
\ee
and $\omega_j$'s are the half periods satisfying $\sum_j\omega_j=0$.
The new coordinates $u_1$ and $u_2$ are related to the old ones by
\be\label{eq:coordinate1}
x=\wp (u_1)+\frac{1+z}{3}\,,\qquad \bar{x}=\wp (u_2)+\frac{1+z}{3}\,,
\ee
which can also be expressed alternatively as
\be\label{eq:coordinate2}
\d u_1=\frac{\d x}{4 y_{x}}\,,\qquad \d u_2=\frac{\d\bar{x}}{4 y_{\bar{x}}}\,.
\ee

Let us make several remarks on the results we got. Firstly the $BC_2$ Inozemtsev model \eqref{eq:hamiltonianino} gives a one-parameter family of differential equations parameterized by the worldsheet cross ratio $z$. In particular, in the limits $z\to 0, 1 ,\infty$,  the elliptic curve \eqref{eq:deftorus} degenerates and the $BC_2$ Inozemtsev model reduces to the $BC_2$ Calogero-Sutherland model. This is consistent with the observations made in \cite{Isachenkov:2016gim,Schomerus:2016epl}, which pointed out the equivalence between standard conformal Casimir differential equations and the $BC_2$ Calogero-Sutherland model. Our result extends their results by unifying Calogero-Sutherland models associated with different OPE channels into a single model. Secondly the $BC_2$ Inozemtsev model is known to be integrable \cite{inozemtsev1989lax,takemura2003inozemtsev}. It would be an interesting future problem to construct a complete basis of eigenfunctions by making use of integrability. Thirdly the parameter ${\sf a}$, which governs the interaction strength between $u_{1,2}$ in \eqref{eq:hamiltonianino}, vanishes in AdS$_3$. In that case, \eqref{eq:hamiltonianino} can be rewritten as two decoupled Heun equations \cite{takemura2003heun}, and the analysis becomes much simpler. Lastly the coordinate transformations \eqref{eq:coordinate1} and \eqref{eq:coordinate2} coincide with the ``pillow'' coordinates for the Virasoro conformal block in 2d CFT, which were introduced first by Zamolodchikov \cite{zamolodchikov1987conformal} and recently revisited in \cite{Maldacena:2015iua} to analyze analytic properties of the four-point functions. To the best of our knowledge, our result is the first example in which the pillow coordinates naturally appear in higher dimensions. It would be interesting to further explore the implications of the pillow coordinates in higher-dimensional CFTs.
\section{\label{sec:discussion}Discussion}
In this paper we constructed a bosonic ambitwistor string theory on a coset manifold. Although anomalies have prevented us from computing amplitudes exactly in the quantum regime of the worldsheet theory, we have applied it to the computation of tree-level amplitudes of bi-adjoint scalar theories in AdS in arbitrary space-time dimension in the classical limit on the worldsheet. The results are given by integrals over the moduli space of punctured Riemann spheres, which localize on an operator-valued version of scattering equations. We then developed a method to evaluate such integrals by making use of a series of contour deformations, and showed that the result agrees with direct perturbation theory in AdS.

Our construction can be viewed as a natural extension of the CHY formalism to the AdS space-time, and potentially provides a useful framework to study scattering amplitudes in AdS, which are dual to correlation functions in strongly-coupled CFTs. Our results constitute a proof-of-principle for this formalism, which needs to be developed in further various directions in order to turn it into a powerful computational tool. We therefore end this paper with a list of future directions.

\paragraph{Extension to other theories.} A natural next step would be to generalize our construction to other theories, in particular to gauge theories and gravity, with and without supersymmetry. Higher-point amplitudes in those theories are much harder to compute from standard perturbation theory in AdS because of a proliferation of Witten diagrams. By contrast, the ambitwistor string treats amplitudes with all multiplicity on a completely equal footing, and would allow us to write down a closed formula for them.  Works in that direction are in progress and we will report their outcome soon. 

\paragraph{Direct evaluation of the AdS scattering equations.} In this paper, we evaluated the integrals over ${\cal M}_{0,n}$ using contour deformations and relating them to Witten diagrams. Although this provided a proof of the equivalence with standard perturbation theory, computationally it is not a real gain.  It would be desirable to develop an alternative approach. One possibility is to decompose the integrand into eigenfunctions of the scattering equations and replace the operator-valued equations with $c$-numbers. This would allow us to evaluate the integrals on the Riemann spheres on the solutions to the scattering equations, in the same way as it can be done for the CHY formalism in flat space, see, e.g., \cite{Cachazo:2013gna,Cachazo:2016ror}.  We took an initial step in this direction in Section \ref{sec:eigenfunctions} by analyzing the eigenvalue equations for the scattering equations for four-point functions, but more works are needed to complete the analysis. It is also worth mentioning that the eigenfunctions of the scattering equations are by themselves interesting objects, since they interpolate between conformal partial waves in different OPE channels. 
\paragraph{Other backgrounds.} Our construction of the ambitwistor string can be applied to more general coset manifolds (although also in that case anomalies will not cancel). It would be interesting to analyze other physically interesting setups, including the de Sitter space \cite{Strominger:2001pn,Witten:2001kn,Maldacena:2002vr,Arkani-Hamed:2015bza} and the holographic duals of non-relativistic CFTs \cite{SchaferNameki:2009xr}. Since the wave-function in the Bunch-Davies vacuum of the de Sitter space is related more or less straightforwardly to the correlation functions in AdS \cite{Maldacena:2002vr,Arkani-Hamed:2015bza}, our formalism would work also in that case. On the other hand, late-time correlation functions in de Sitter are harder to analyze since one has to perform the path integral along the Schwinger-Keldysh contour, which would amount to preparing two copies of the de Sitter spaces and gluing them together at late time.

\paragraph{Flat space limit.} Our formula for the correlation functions resembles in many respects the CHY formula for the flat-space S-matrix. At a formal level, one can arrive at our  formula by replacing each factor in the integrand of the CHY formula with its AdS counterpart; for instance $p_i\cdot p_j \mapsto D_i\cdot D_j $. This however does not immediately guarantee that the flat-space limit of our formula reproduces the CHY formula. This is because taking the flat-space limit of the correlation functions in AdS involves integrals of the boundary points, or alternatively the integrals in the Mellin space, as was discussed in \cite{Penedones:2010ue, Okuda:2010ym}. It would be important to work out the details of the flat-space limit.  This might shed light on the soft theorem in flat space; see recent discussions in \cite{Hijano:2020szl}.

\paragraph{Loops.} In flat space, the CHY formalism and the ambitwistor string were extended also to loop amplitudes \cite{Adamo:2013tsa,Geyer:2015bja,Geyer:2015jch,Geyer:2016wjx,Roehrig:2017gbt,Geyer:2017ela,Geyer:2018xwu,Geyer:2019hnn}. Unfortunately, one cannot follow the same path in the present case since the theory is anomalous at the quantum level. We therefore need to come up with an alternative approach. One possibility is to make use of recent progress in the application of unitarity methods to the correlation functions in AdS \cite{Meltzer:2019nbs}. It would be interesting to see if one can apply such methods to our CHY-like representation.

\paragraph{Double-copy and the relation to sectorized string.}

One of the advantages of the CHY formalism in flat space is that it manifests various double-copy relations between amplitudes in gauge and gravity theories \cite{Cachazo:2013iea,Cachazo:2014xea}, which are completely obscured from the Feynman-diagram point of view. One may hope that a similar structure might extend to AdS space (see related work \cite{Farrow:2018yni,Lipstein:2019mpu} for $3$-point correlators in momentum space). However, our construction reveals an interesting obstruction to the naive version of double copy in AdS. In the notation of \eqref{eq:intro}, the expectation is that gauge and gravity theory AdS amplitudes will be computed with some operators ${\cal I}_\text{L}$ and ${\cal I}_\text{R}$, which in general might not commute. One would have to account for this fact in a modified version of double-copy. Making this statement more precise will require a computation of the appropriate integrands coming from spin-$1$ and $2$ vertex operators, which will be given in a future publication. Finally, it might be interesting to construct a version of our AdS model in the context of sectorized/chiral string \cite{Siegel:2015axg,Jusinskas:2016qjd,Casali:2016atr,Azevedo:2017yjy,Lee:2017utr,Azevedo:2019zbn}, together with its intersection theory interpretation \cite{Mizera:2019gea}, which introduces a non-trivial $\alpha'$ dependence to double-copy.

\section*{Acknowledgement}
We thank Eduardo Casali, Zohar Komargodski, Petr Kravchuk, Juan Maldacena, Kai Roehrig, David Skinner, and Edward Witten for useful conversations. We also thank Edward Witten for pointing out incorrect statements made in the first version of this paper regarding the ambitwistor string on a coset manifold and its anomalies.
 LE gratefully acknowledges support from the Della Pietra family at IAS.  The work of SK is supported by DOE grant number DE-SC0009988. SM gratefully acknowledges the funding provided by Carl P. Feinberg.

\appendix

\section{\label{sec:notation}Summary of the notation}

Throughout the paper we need many different kinds of symbols. In order to improve readability, the most used notation is summarized here.
\paragraph{Indices.}
\begin{itemize}
\item $a$, $b$, $c$, \dots: adjoint indices of the group $\text{G}_\text{L}$, where $\text{G}$ is the denominator group of the coset. We mostly take $\text{G}=\mathrm{SO}(d,2) \times \mathrm{SO}(d+2)$.
\item $\bar{a}$, $\bar{b}$, $\bar{c}$, \dots: adjoint indices of $\text{G}_\text{R}$.
\item $r$, $s$, $t$, \dots: adjoint indices of the gauged subgroup $\mathrm{H} \subset \text{G}_\text{R}$.
\item $i$, $j$, $k$, \dots: particle labels.
\item $A$, $B$, $C$, \dots: embedding space index. The embedding space formalism is reviewed in Section~\ref{sec:embedding-space}.
\item $\alpha$, $\beta$, $\gamma$, \dots: adjoint indices of the first gauge group of the bi-adjoint scalar.
\item $\tilde{\alpha}$, $\tilde{\beta}$, $\tilde{\gamma}$, \dots: adjoint indices of the second gauge group of the bi-adjoint scalar.
\end{itemize}
\paragraph{Worldsheet fields.}
\begin{itemize}
\item $g$: the group-valued field of the ambitwistor string.
\item $p$: the analogue of $P$ in the standard flat-space ambitwistor string. It takes valued in the cotangent bundles $T^*\text{G}$.
\item $T$: the worldsheet energy-momentum tensor.
\item $H$: the second spin-2 field on the worldsheet that gauges the light-cone rescaling. It is the analogue of $P^2$ in the flat-space ambitwistor string.
\item $K^\alpha$ and $K^{\tilde{\alpha}}$: the generators of the two internal current algebras on the worldsheet.
\item $b$, $c$: worldsheet ghosts that gauge $T$.
\item $\tilde{b}$, $\tilde{c}$: worldsheet ghosts that gauge $H$.
\item $b_r$, $c_r$: worldsheet ghosts that gauge $J_\text{R}^r$ and reduce the model to the coset.
\item $Q$: the worldsheet BRST operator.
\item $R$: Vector in $\mathds{R}^{d,2}$, whose stabilizer inside $\mathrm{SO}(d,2)$ is the subgroup we are gauging. It satisfies $R \cdot R=-1$.
\end{itemize}
\paragraph{Embedding space.}
\begin{itemize}
\item $P$: boundary embedding space coordinate. Satisfies $P^2=0$ and $P \sim \lambda P$ for $\lambda \in \mathds{R}\setminus\{0\}$. Thus expressions are always homogeneous in $P$.
\item $X$: bulk embedding space/$\text{AdS}_{d+1}$ coordinate. Satisfies $X \cdot X=-1$. In the worldsheet theory, this is a field that is defined as $X=gR$.
\item $D^a=D^{AB}$: the conformal generators in the embedding space.
\item $\Box_{X}$: $\text{AdS}_{d+1}$ Laplacian.
\end{itemize}
\paragraph{CHY formula.}
\begin{itemize}
\item $E_i$: scattering equations, see eq.~\eqref{eq:scattering-equations}.
\item $\mathcal{C}=\mathcal{C}(P_1,\dots,P_n)$: the contact diagram in $\text{AdS}_{d+1}$ (D-function) \cite{Freedman:1998tz}.
\item $m_\text{AdS}$: bi-adjoint scalar $\text{AdS}_{d+1}$ amplitude.
\item $m_\text{AdS}(\sigma|\tau)$: doubly color-ordered bi-adjoint scalar $\text{AdS}_{d+1}$ amplitude.
\item $\hat{m}_\text{AdS}(\sigma|\tau)$: doubly color-ordered bi-adjoint scalar $\text{AdS}_{d+1}$ amplitude with contact diagram stripped off.
\end{itemize}

\bibliographystyle{JHEP}
\bibliography{bib}
\end{document}